\documentclass[twocolumn,floats,aps,pre,superscriptaddress,showpacs,floatfix]{revtex4-1}

\usepackage[colorlinks=true, citecolor=cyan]{hyperref}

\usepackage{amsmath,amssymb,amsfonts,color,graphicx}

\newcommand{\K}{\operatornamewithlimits{\text{\LARGE K}}}
\newcommand{\wt}[1]{\widetilde{#1}}

\begin{document}

\title{Kinetics of self-assembly via facilitated diffusion: formation of the transcription complex}
\begin{abstract}
We present an analytically solvable model for self-assembly of a molecular complex on a filament. The process is driven by a seed molecule that undergoes facilitated diffusion, which is a search strategy that combines diffusion in three-dimensions and one-dimension. Our study is motivated by single molecule level observations revealing the dynamics of transcription factors that bind to the DNA at early stages of transcription. We calculate the probability that a complex made up of a given number of molecules is completely formed, as well as the distribution of completion times, upon the binding of a seed molecule at a target site on the filament (without explicitly modeling the three-dimensional diffusion that precedes binding). We compare two different mechanisms of assembly where molecules bind in sequential and random order. Our results indicate that while the probability of completion is greater for random binding, the completion time scales exponentially with the size of the complex; in contrast, it scales as a power-law or slower for sequential binding, asymptotically. Furthermore, we provide model predictions for the dissociation and residence times of the seed molecule, which are observables accessible in single molecule tracking experiments.
\end{abstract}


\author{Ziya Kalay}
\affiliation{Institute for Integrated Cell-Material Sciences (WPI-iCeMS), Kyoto University, Yoshida Ushinomiya-cho, 606-8501, Kyoto, Japan}

\date{\today}
\maketitle

\section{Introduction}
\label{sec:introduction}
Many biochemical processes involve formation of mesoscopic molecular structures that perform complex tasks. One well-known example is the transcription complex which plays the key role in accessing the information coded in the DNA~\cite{alberts_molecular_2007}. During the process of transcription, a complex consisting of RNA polymerase and transcription factors is assembled on the DNA in order to \emph{read} the genetic information, and produce RNA molecules. Thanks to powerful methods of molecular biology, it has been possible to study the number and types of molecules involved in the formation of the transcription complex; nevertheless, kinetics of the formation of transcription complex is much less known, and is now an active area of biophysics~\cite{stasevich_assembly_2011,mueller_quantifying_2013,coulon_eukaryotic_2013}. A key experimental finding~\cite{hammar_lac_2012} regarding the kinetics of transcription factors is that at least some of the molecules that bind to sites on DNA undergo \emph{facilitated diffusion}~\cite{berg_diffusion-driven_1981,von_hippel_facilitated_1989}, during which molecules diffusing in three-dimensions (3-D) can temporarily bind to the DNA, diffusing along the filament and densely exploring it, which is thought to be an efficient search strategy~\cite{hammar_lac_2012}. Other aspects of faciliated diffusion, such as how it would affect the noise in transcriptional regulation, have also been explored in previous theoretical work~\cite{tkacik_diffusion_2009, paijmans_lower_2014}.   
\begin{figure}[t]
\centering
\includegraphics[width=0.95\columnwidth]{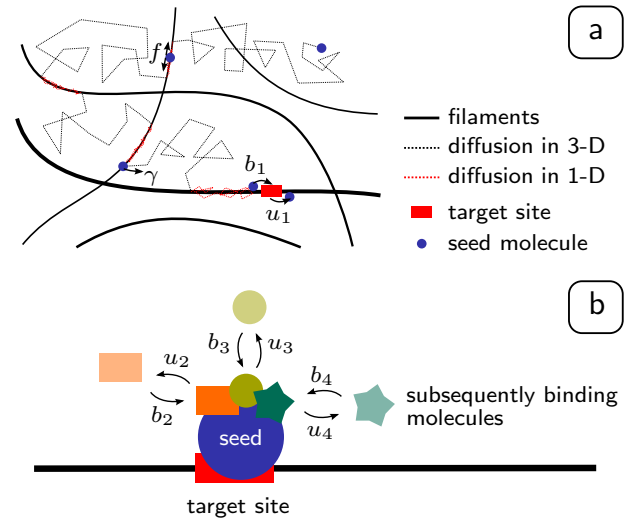}
\caption{(Color online) Illustration of the process of self assembly via facilitated diffusion. In (a), facilitated diffusion of the seed molecule is shown, where it performs Brownian motion in 3-D (not explicitly modeled in this work) and temporarily binds to filaments, diffusing in 1-D (at a rate $\propto f$). When the seed arrives at the target site, it binds at rate $b_1$, and if it is already bound, becomes unbound at rate $u_1$. In (b), assembly process is shown for $w=4$. After the seed is bound, additional molecules are recruited via reversible binding.}
\label{fig:illustration-biological}
\end{figure}

We consider the kinetics of a self-assembly process in which a \emph{seed} molecule diffuses in 3-D, and gets temporarily attached to a filament that carries a target site (see Fig. \ref{fig:illustration-biological} for an illustration of the process and of facilitated diffusion). In this work, we do not explicitly model diffusion in 3-D, which has been studied earlier~\cite{berg_diffusion-driven_1981}, and focus on the dynamics of a nucleation process initiated by the seed, as described below. While it is associated with the filament, the seed molecule undergoes one-dimensional (1-D) diffusion and when it occupies the target site, it becomes bound at a constant rate, triggering the subsequent binding of other molecules that bind and unbind at constant rates. When $w$ molecules are assembled, the process is complete. We consider the case where the seed molecule is initially bound at the target site, and focus on the kinetics of the rest of the process. Behavior of the seed molecule is motivated by the observation of facilitated diffusion of transcription factors in bacterial cells, as mentioned above. We envisage that the seed is an essential molecule for transcription initiation which possesses binding sites for other molecules or induces the binding of additional transcription factors, such as the RNA polymerase~\cite{alberts_molecular_2007, stasevich_assembly_2011, mueller_quantifying_2013, coulon_eukaryotic_2013}. In eucaryotic cells, RNA polymerase II (RNAp2) transcribes the majority of genes. Although the binding order of molecules that form the transcription initiation complex is not clear, experiments suggest that a number of transcription factors need to bind both before and after the binding of RNAp2~\cite{coulon_eukaryotic_2013}. Therefore, if one were to think of RNAp2 as the seed molecule, the model presented here would be applicable to the latter part of the assembly process, starting with the binding of RNAp2. Alternatively, the seed can represent a molecule that binds during the initial stages of assembly. A candidate for such a molecule is TFIID, which significantly changes the local conformation of the DNA, paving the way for subsequent molecules to bind~\cite{alberts_molecular_2007}.

Our main results consist of an exact expression for the probability that the assembly completely forms upon the binding of a seed molecule, and the distribution of the completion time. In addition to these quantities that describe the kinetics of assembly, and to link model predictions with quantities that can be accessed in single molecule observations, distributions of the time at which the seed dissociates from the filament, as well as the residence time in an interval containing the target site are also presented. 

In what follows, we first describe a mathematical model corresponding to the process described above and illustrated in Fig. \ref{fig:illustration-biological}. We then present results of analytic calculations for the quantities mentioned above. Lastly, we discuss the applicability of the model as well as previous findings on the same problem, and state our conclusions. We also provide a comparison of analytical results with simulations of the process, presented in Appendix \ref{sec:simulation}.

\section{An Analytically Solvable Model of Assembly}
\label{sec:model-description}
\subsection{Model description and assumptions}
\begin{figure}[t]
\centering
\includegraphics[scale=1]{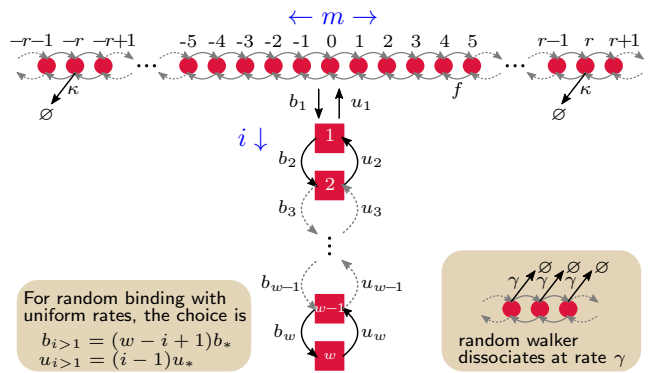}
\caption{(Color online) Illustration of the assembly model as a Markov chain. Circles and squares correspond to different states of the system, and arrows show state transitions with corresponding rates. Circles denote different positions on the filament, indexed by $m$, modeled as a lattice, where the seed molecule performs a random walk between adjacent sites, with a hopping rate $f$. As shown in the lower right, the random walker disappears from the system at a constant rate $\gamma$ only while it is diffusing along the filament [states indicated by circles]. Squares correspond to the bound states indexed by $i$. For sequential and random binding models, the rates $b_{i>1}$ and $u_{i>1}$ are interpreted in a different fashion (see text for explanation).}
\label{fig:illustration-Markov-chain}
\end{figure}

We consider the formation of a molecular complex consisting of $w$ elements that are assembled sequentially or in random order. Formation of the complex is initiated by a \emph{seed molecule}, which we will just refer to as the \emph{seed}. The seed is envisaged to diffuse in 3-D (not explicitly modeled) and temporarily attaches to filaments along which it undergoes 1-D diffusion. When the seed occupies the target site on the filament, it can become bound at rate $b_{1}$. 

After the seed is bound, a total number of  $w-1$ molecules start to assemble, which are, so to speak, recruited by the seed. If the seed is the only molecule in the complex, it becomes unbound at rate $u_{1}$, returning to diffusion along the filament. When two or more molecules (including the seed) are bound at the same time, the seed cannot become unbound. Note that this assumption may not be applicable in general, and the model considered in this work is appropriate for the case where the binding of subsequent molecules stabilizes the complex. While the seed is unbound and diffusing on a filament, it dissociates at a constant rate $\gamma$. When $w$ molecules are assembled, the process is complete.

In this work, we do not explicitly model the motion of the seed in 3-D, and focus on the dynamics when the seed molecule is initially bound at the target site. This approach allows us to study the kinetics of the assembly process in the presence of a low concentration of seed molecules, which is often a good assumption in cell biology~\cite{phillips_physical_2008}, accounting for the effect of facilitated diffusion. For a treatment of the problem of searching for a target site in a filament via 3-D diffusion interrupted by periods of 1-D exploration, we refer the readers to existing literature~\cite{von_hippel_facilitated_1989, tkacik_diffusion_2009, meyer_geometry-induced_2012, paijmans_lower_2014, kaizu_berg-purcell_2014}.  

While a cell can contain a large number of transcription factors as well as corresponding target sites, we consider a regime where the concentration of seed molecules is low such that the competition for the target site is negligible, and the dynamics of the system is well characterized by a single seed and a target site. Nevertheless, we do provide a generalization for multiple molecules under the assumption of negligible competition (see Section \ref{sec:multiple-seeds}).

\subsection{Mathematical details and predicted quantities}
Based on the biologically inspired model illustrated in Fig. \ref{fig:illustration-biological}, we consider a model for diffusion and assembly that can be analytically solved by standard tools of statistical mechanics~\cite{krapivsky_kinetic_2010}. See Fig. \ref{fig:illustration-Markov-chain} for an illustration of the corresponding mathematical model.

Diffusion along the filament is modeled as a continuous time random walk in a 1-D lattice, where the random walker, that is the seed, hops between adjacent lattice sites at a (symmetric) rate $f$ (see Fig. \ref{fig:illustration-Markov-chain}). For a DNA filament, it is natural to think that the lattice sites correspond to base pairs that are separated by $\approx 0.34$ nm~\cite{alberts_molecular_2007}. The lattice site with index $m=0$ is where the target is located. When the seed is occupying site $m=0$, it can become bound at rate $b_{1}$, upon which the system would transition to the first bound state $i=1$. Being in the first bound state, the system can either go back to the state where the seed is diffusing, at rate $u_{1}$, or transition to the next bound state, $i=2$, if an auxiliary molecule becomes bound. Note that while the meaning of the index $m$ is straightforward (the position of the seed), the physical picture ascribed to the $i^{\rm th}$ bound state depends on the details of how molecules are assembled. We consider two different models of assembly: sequential and random.

In sequential assembly, $w-1$ auxiliary molecules can reversibly bind in order, once the seed becomes bound. When the $i^{\rm th}$ auxiliary molecule binds or unbinds, the system transitions to the $i+1^{\rm st}$ or $i^{\rm th}$ bound state, respectively. The binding and unbinding rates for the $i^{\rm th}$ auxiliary molecule is equal to $b_{i+1}$ and $u_{\rm i+1}$, respectively.

If auxiliary molecules assemble in random order, equally likely in any of the $(w-1)!$ possible ways, the model illustrated in Fig. \ref{fig:illustration-Markov-chain} can still be used, provided that the binding and unbinding rates of different auxiliary molecules are sufficiently similar. We suppose that each molecule binds and unbinds at the uniform rates $b_{*}$ and $u_{*}$, respectively. Provided that this is the kinetics at the level of individual molecules, the transition rates for the whole system would become $b_{i>1} = (w-i+1)b_{*}$ and $u_{i>1} = (i-1)u_{*}$, which was also employed in a previous study~\cite{dorsogna_first_2005}. This is a consequence of treating the reactions involving auxiliary molecules as Poisson processes with exponential waiting times. When $j$ of the auxiliary molecules are bound, such that the system is in state $i=j+1$, binding of any of the remaining $w-1-j$ molecules would take the system to the $j+2^{\rm nd}$ state, or unbinding of any of the $j$ molecules would take the system to the $j^{\rm th}$ state. The former takes place at rate $(w-j)b_{*}$, whereas the latter at $ju_{*}$, since the minimum of a set of independent exponential random variables is also distributed exponentially, with a parameter that equals the sum of all individual random variables' parameters.

In addition to studying the kinetics of the assembly, we are also interested in making predictions for observables relevant in single-molecule tracking experiments. Supposing that the seed is labeled (via fluorescent dyes, quantum dots, etc.) and its position can be tracked, one may be able to observe the time it takes for the seed to dissociate from the filament, or the time it takes for it to exit from an interval of length $2r$ centered at the target site, given the seed was observed to be bound at $t=0$. To be able to calculate the statistics of these two times from the model, we introduce, for solely mathematical convenience, two \emph{leaky} sites at $m=-r$ and $r$, where the seed disappears from the system at rate $\kappa$. In the next section, we consider the limits $\kappa \to 0$ and $\kappa \to \infty$ depending on which calculation is of concern. 

Based on the model illustrated in Fig. \ref{fig:illustration-Markov-chain}, we write a set of master equations for the probability of finding the seed (unbound) at lattice site $m$, denoted by $P_{m}(t)$, and the probability of finding the system in the $i^{\rm th}$ bound state, denoted by $Q_{i}(t)$, at time $t$. Note that $Q_w(t)$ is the probability of having a fully assembled complex at time $t$, from which we will derive the probability of completion as well as the first completion time, in the next section. The master equation as well as its analytical solution is given in Appendix \ref{sec:master-eqn-solution}. In the next section, we present results derived from the probabilities $P_{m}$ and $Q_{i}$, assuming that they are known, and always refer to appendices for calculation details. 

\section{Results}

In this section we present results for the kinetics of the assembly process obtained by solving the model illustrated in Fig. ~\ref{fig:illustration-Markov-chain}. Results are displayed in a way that highlights the difference between the kinetics for sequential and random binding models. For convenience, and to be able to treat the case of random binding as described in the previous section, we consider uniform rates $b_{*}$ and $u_{*}$ as the binding and unbinding rates of each auxiliary molecule, regardless of order.

All rates and times are measured in units of $f$ and $1/f$, respectively, where $f$ is the hop rate between adjacent sites in the lattice, proportional to the 1-D diffusion coefficient.

\subsection{Probability of completion}
A key quantity that characterizes the efficiency of the assembly process is the probability that the molecular complex completely forms before the seed dissociates from the filament, given the seed was initially bound (that is, $Q_{1}(0)=1$, $Q_{i\neq 1}(0)=0$ and $P_m(0)=0$). We refer to this quantity as the probability of completion, and denote it by $P_{\rm comp}$. We note that $P_{\rm comp}$ is the probability of arriving at the bound state $w$ at any time as $t \to \infty$, which is also the fraction of system trajectories that end at $w$. Therefore, we have $P_{\rm comp} = \lim_{t \to \infty} Q_{w}(t)$, under the condition $u_{w}=0$, ensuring that trajectories that reach the last bound state are terminated. Performing the calculation, we obtain (see Appendix \ref{sec:PcompTcomp})
\begin{gather}
P_{\rm comp} = \frac{ 1 }{ 1 + \displaystyle\frac{\lambda(w)}{1 + \beta} },
\label{eqn:Pcomp}
\end{gather}
where $\beta = b_{1}/\sqrt{\gamma(\gamma + 4f)}$ and $\lambda(w)$ is a constant formed by the combination of all the rates $b_{i}$ and $u_{i}$ except $b_{1}$, explicitly given in \eqref{eqn:lambda-w}. If the complex is made up of just a pair of molecules, $\lambda(w)$ has a particularly simple form, given by $\lambda(2) = u_1$. Note that $\beta$ quantifies the ratio of the affinity to the binding site (i.e., nucleation rate given the particle occupies the binding site) to the rate at which the seed is carried away from the binding site, either via dissociation or diffusion. When $f/\gamma \ll 1$, meaning dissociation is much more rapid than diffusion along the filament, we have $\beta \propto b_{1}/\gamma$; and when $f/\gamma \gg 1$, meaning diffusion is much faster than dissociation, $\beta \propto b_{1}/\sqrt{f \gamma}$. 

\begin{figure}[t]
\centering
\includegraphics{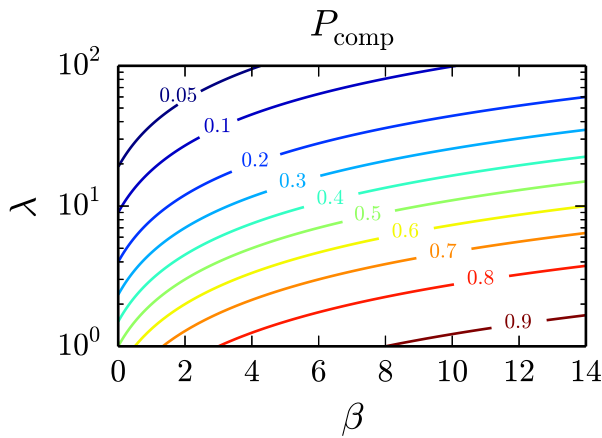}
\caption{(Color online) Contour plot of the probability of completion as a function of $\lambda$ and $\beta$.}
\label{fig:Pcomp-contour}
\end{figure}
Since $\lambda$ and $\beta$ do not share any parameters, they constitute a good pair of knobs that can be used to investigate the behavior of $P_{\rm comp}$. In Fig.~\ref{fig:Pcomp-contour}, a contour plot of $P_{\rm comp}$ is displayed as a function of $\lambda$ and $\beta$. We note that it becomes less and less probable for the complex to be completed as $\lambda$ increases, or $\beta$ decreases. This is in line with what one may expect intuitionally, since larger $\lambda$ values correspond to relatively larger unbinding rates $u_{i>1}$, and smaller $\beta$ values imply that the seed is diffusing fast and/or it dissociates from the filament quickly.

As $\lambda$ and $\beta$ are combinations of many parameters, it is informative to explore the behavior of $P_{\rm comp}$ as a function of parameters whose physical meaning is more direct. In this respect, next, we display how $P_{\rm comp}$ changes with the total number of molecules in the complex and the ratio of the binding and unbinding rates $b_*/u_*$.

\begin{figure}[t]
\centering
\includegraphics{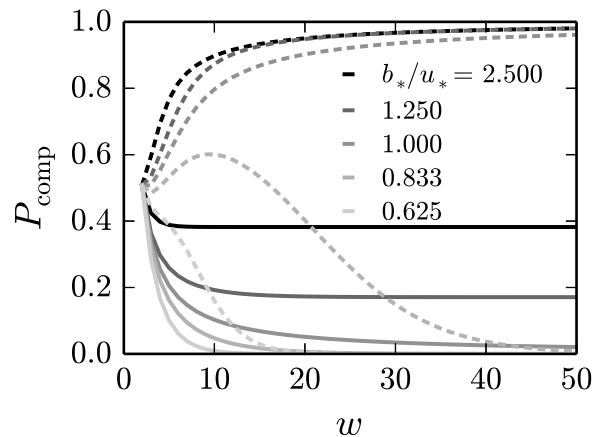}
\caption{$P_{\rm comp}$ as a function of the total number of molecules in the complex for different values of the ratio $b_*/u_*$ (obtained by varying $u_*$). Solid and dashed curves correspond to sequential and random binding, respectively. Parameter values are: $\gamma=0.1$, $b_1=2$, $b_*=0.25$, and $u_1=1$, measured in units of $f$, the hopping rate along the filament.}
\label{fig:Pcomp-vs-w}
\end{figure}

In Fig.~\ref{fig:Pcomp-vs-w}, $P_{\rm comp}$ is plotted as a function of $w$, for five different values of $b_{*}/u_{*}$, indicated by curves with different shades of gray. In this figure and throughout the article, dashed and solid curves correspond to random and sequential binding models, respectively, unless otherwise noted. We see that $P_{\rm comp}$ has greater values for random binding compared to sequential binding for the same set of parameter values. In random binding, $P_{\rm comp}$ can be non-monotonic in the number of bound states, depending on the value of the ratio $b_*/u_*$. 


The behavior of $P_{\rm comp}$ as a function of $w$ can be intuitively understood, to a certain extent, by ignoring the diffusion states and associating $m=0$ with dissociation (see Fig. \ref{fig:illustration-Markov-chain}). Then the problem can be viewed as (biased) random walk in a 1-D lattice with $w+1$ sites, $i=0,\,1,\,...,\,w$ where arriving at the top (site 0) means dissociation, and arriving at the bottom (site $w$) means completion, and the random walker starts at site $i=1$. We take $b_1 = b_*$ for simplicity. In the next three paragraphs, we provide an intuitive explanation for the behavior observed in Fig. \ref{fig:Pcomp-vs-w}.

In sequential binding, the rate of acquiring and losing an auxiliary molecule does not depend on the number of already bound molecules. When $b_*/u_*>1$, the random walk is biased downward, leading to the completion of the process with a probability that increases with $b_*/u_*$ (see Fig. \ref{fig:Pcomp-vs-w}). On the contrary, for $b_*/u_*<1$, the bias is against any motion towards completion. Therefore, the only way for the random walker to end up at the bottom first is via a highly improbable sequence of downward steps, whose probability is expected to dramatically diminish with increasing $w$, leading $P_{\rm comp}$ to zero as a function of $w$. 

In random binding, when $n$ molecules are bound, the rate at which the complex grows is given by $(w-n)b_*$, and the rate at which it shrinks is $nu_*$. Therefore, there is a state with $n_*=b_*w/(u_*+b_*)$ bound molecules, which is more stable than others in the sense that the growth and shrinking rates are approximately balanced. Note that we have $b_*/u_* \approx n_*/(w-n_*)$, implying that for $b_*/u_*>1$ we expect to have, \emph{on average}, more bound molecules than missing ones. On the other hand, for $b_*/u_*<1$, we expect to have more missing molecules. When $b_*/u_*>1$, the stable state is in the lower half of the simplified lattice, and fluctuations are more likely to drive the system to the bottom, i.e. completion. In the other case, $b_*/u_* < 1$, the stable state is in the upper half, and fluctuations are more likely to take the system to the top, i.e. dissociation. This intuitive picture is in line with the results shown in Fig. \ref{fig:Pcomp-vs-w}. Note that the the curve that corresponds to $b_*/u_*=0.833$ indicates that the stable state in this case moves from the lower half to the upper half at around $w\approx 15$.

In the random binding model, when $b_*/u_*<1$, we observe a transient increase in $P_{\rm comp}$, although it eventually decays to zero with increasing $w$. To understand this, we note that the completion probability can be expressed by $(1-p)q$, where $p$ is the probability of dissociation during the first transition, and $q$ is the probability of completion starting from site 2 (as the random walker of the simplified problem would end up in state 2 if it did not dissociate after the first step). Note that $p = 1/(1+(w-1)b_*/u_*)$, the probability of taking the first step upward, decreases with $w$ (the number of auxiliary molecules). Therefore, one would expect an initial increase in $P_{\rm comp}$ due to the increasing factor $(1-p)$. Nevertheless, for $b_*/u_*<1$, this increase is balanced by a decrease in $q$, as completion before dissociation gets more and more improbable as the number of steps required to arrive at $w$ increases, as discussed in the previous paragraph. 

On the whole, we find that there is a qualitative difference in the behavior of $P_{\rm comp}$ for random and sequential binding, and that there can be an optimal value of $w$ that maximizes $P_{\rm comp}$ in random binding, depending on the relative strength of binding and unbinding rates of the auxiliary molecules.

\begin{figure}[b]
\centering
\includegraphics{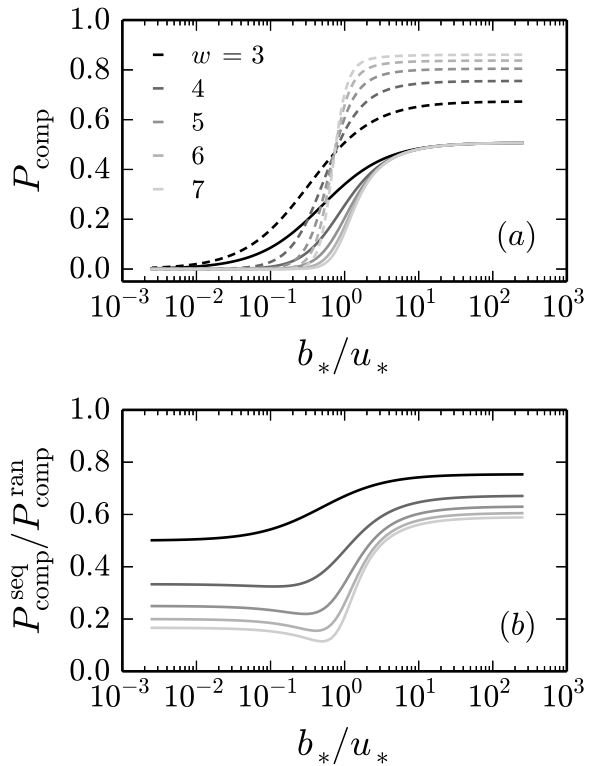}
\caption{Probability of completion as a function of the ratio $b_*/u_*$. In (a), $P_{\rm comp}$ is plotted for different values of the total number of molecules in the complex. Solid and dashed curves correspond to sequential and random binding, respectively. In (b), ratio of $P_{\rm comp}$ for sequential binding to that for random binding is displayed. Parameter values for both graphs are: $\gamma=0.1$, $b_1=1$, $b_*=0.25$, and $u_1=1$, measured in units of $f$, the hopping rate along the filament.}
\label{fig:Pcomp-vs-u}
\end{figure}

Next, we consider how the probability of completion depends on the ratio $b_{*}/u_{*}$ for different values of the total number of molecules. Fig.~\ref{fig:Pcomp-vs-u} (a) shows $P_{\rm comp}$ as a function of $b_{*}/u_{*}$ for $3 \le w \le 7$. We note that $P_{\rm comp}$ monotonically increases with $b_{*}/u_{*}$ for both random and sequential binding models. In sequential binding, $P_{\rm comp}$ does not depend on $w$ for $b_*/u_* \gg 1$. In contrast, the values of $P_{\rm comp}$ at $b_{*}/u_{*} \gg 1$ increase with $w$, by as much as $\approx 25\%$ in the random binding model. In Fig.~\ref{fig:Pcomp-vs-u} (b), the ratio $P_{\rm comp}^{\rm seq}/P_{\rm comp}^{\rm ran}$, where the superscripts denote the binding order of auxiliary molecules, is plotted as a function of $b_{*}/u_{*}$. We observe that $P_{\rm comp}$ is greater for the random binding model, especially for $b_{*}/u_{*} > 1$, and that there is a large drop in the ratio around $b_{*}/u_{*}=1$.

Lastly, we present results quantifying the effect of facilitated diffusion on how likely the process is going to be completed. Given the seed is initially bound with no other bound molecules, setting $b_1=0$ would prevent the possibility of rebinding since the seed cannot become bound again if it ever transits to the diffusive state. Therefore, the ratio $\rho = P_{\rm comp}/P_{\rm comp}(b_1=0)$ would give us the relative enhancement of completion probability due to facilitated diffusion. Note that setting $f=0$ maximizes the enhancement due to rebinding, as in this case a seed that becomes unbound does not leave the binding site ($m=0$) via 1-D diffusion, such that rebinding occurs at the maximal rate $b_1$.

The relative enhancement ratio $\rho$ is explicitly given by
\begin{gather}
\rho = \frac{ 1 + \lambda }{ 1 + \displaystyle\frac{\lambda}{ 1 + \beta } }.
\label{eqn:fast-rebinding-efficiency}
\end{gather}
Expanding \eqref{eqn:fast-rebinding-efficiency} around $\beta=0$, we get
\begin{gather*}
\rho = 1 + \frac{ \lambda}{1+\lambda} \beta + O\left( \beta^2 \right),
\end{gather*}
which clearly shows that there is no enhancement when the affinity to binding site is zero, $\beta=0$, or when dissociation or diffusion rates diverge, that is, $\gamma \to \infty$ or $f \to \infty$, implying $\beta \to 0$. Note that the enhancement initially increases linearly with the ratio $\beta$ [defined below \eqref{eqn:Pcomp}]. 

\begin{figure}[t]
\centering
\includegraphics{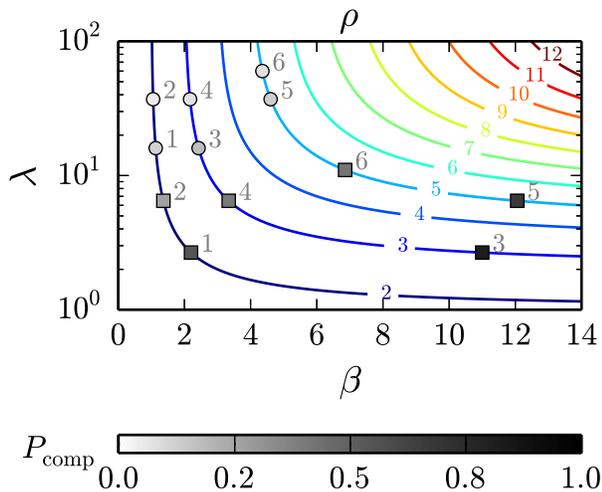}
\caption{(Color online) Contour plot of the enhancement ratio due to facilitated diffusion. Sequential and random binding models are indicated by circles and squares, respectively, and symbols with the same number have identical binding and unbinding rates per molecule: $u_*=0.25$ for 1 and 3; 0.4 for 2, 4 and 5; 0.5 for 6. Points on the same curve correspond to the same enhancement factor for different values of $u_*$, where other parameters are fixed as $w=5$, $\gamma=0.1$, $b_1=2$, $b_*=0.25$, and $u_1=1$, where rates are measured in units of $f$. Data points are color coded in grayscale according to their $P_{\rm comp}$ values and the corresponding color bar is indicated below the graph.}
\label{fig:rho-contour}
\end{figure}

To demonstrate the behavior of the enhancement factor, a contour plot of $\rho$ is displayed in Fig. \ref{fig:rho-contour} as a function of $\lambda$ and $\beta$, plotted in the same domain as $P_{\rm comp}$ in Fig. \ref{fig:Pcomp-contour}. We see that facilitated diffusion enhances the probability of completion as $\lambda$ and $\beta$ increase. Greater $\beta$ values correspond to a situation where the seed spends greater amount of time bound to the filament, in the vicinity of a lattice site, and hence explains the increase in $\rho$. While the parameter $\lambda$ depends on many model parameters, it is roughly proportional to $u_{*}/b_{*}$. Therefore, greater $\lambda$ values correspond to relatively larger unbinding rates, which hinders the completion of the process both with and without diffusion. Higher $\rho$ values for increasing $\lambda$ suggest that facilitated diffusion can counter this hindrance effect to a limited extent.

In Fig. \ref{fig:rho-contour}, we also include a number of data points to compare the effect of facilitated diffusion in random and sequential binding. Sequential and random binding models are indicated with circles and squares, respectively. Data points with the same index correspond to the same value of $b_{*}$ and $u_{*}$, and hence the same per molecule binding/unbinding rates for auxiliary molecules, while the rest of the parameters are determined by imposing the condition that $\rho$ is fixed. Each data point is colored in grayscale according to its $P_{\rm comp}$ value, and the color bar is shown at the bottom of the graph. Comparing the relative positions of the data points with the same index and considering the corresponding $P_{\rm comp}$ values, we conclude that with the same kinetic constants per reaction ($b_{*}$ and $u_{*}$), random binding has larger probability of completion; however, achieving the same enhancement factor with random binding requires stronger association with the filament as compared to sequential binding (compare symbols located on the same line).

\subsection{Completion Time}
In the previous section we demonstrated the behavior of the completion probability as a function of a subset of the model parameters and found that the order in which auxiliary molecules bind has a significant effect on the chance of completion, random binding being more efficient. Next, we present model predictions regarding the time-dependence of the process, providing the complementary information to answer the question: given the process completes, how long does it take?

We define $T_{\rm comp}$ as the completion time, the time at which all $w$ molecules are assembled for the first time, given the seed is the only molecule initially bound. Let $f_{\rm comp}(t)$ be the probability density function of the random variable $T_{\rm comp}$, conditioned on the completion of the process. In terms of $Q_{i}(t)$, the conditional distribution $f_{\rm comp}(t)$ can be expressed as
\begin{align*}
f_{\rm comp}(t) &= \frac{1}{P_{\rm comp}} \left\{ -\frac{d}{dt}\left[ 1 - Q_{w} (t;\, u_w=0) \right] \right\}, \\
& =\frac{1}{P_{\rm comp}} \left. \frac{d Q_{w}}{dt} \right|_{u_w=0},
\end{align*}
where the quantity in square brackets in the first line is the survival probability, that is, the probability of not having visited the $w^{\rm th}$ bound state until time $t$, and we need the normalization constant $P_{\rm comp}$, as $f_{\rm comp}$ is conditioned on the completion of the process. In Appendix \ref{sec:PcompTcomp}, we show that the Laplace transform of $f_{\rm comp}(t)$ is given by
\begin{align}
& \widetilde{f}_{\rm comp}(\epsilon) = \Bigg \{ \frac{(-1)^{w-1}}{P_{\text{comp}}} \left( \prod _{i=2}^w K_i(w)  \right) \nonumber \\  & \times \left. \left( \frac{\epsilon}{\epsilon + b_2 + u_1 \alpha(\epsilon)/\left( \alpha(\epsilon)+b_1 \right) + u_2 K_2(w)} \right) \Bigg\} \right|_{u_w = 0},
\label{eqn:fcomp-laplace}
\end{align}
where $\alpha(\epsilon) = \sqrt{(\gamma +\epsilon ) (\gamma +4 f+\epsilon )}$, $K_{i}(w)$ is a continued fraction involving the rate constants [see \eqref{eqn:K-i}], and the Laplace transform is defined as
\begin{gather}
\wt{f}(\epsilon) = \int_{0}^{\infty} dt\, e^{-\epsilon t} f(t).
\label{eqn:laplace-transform-definition}
\end{gather}
Throughout the text, we will use tildes to distinguish Laplace transformed quantities. 

In the following, we first discuss how the mean and variance of $T_{\rm comp}$ depend on the binding order and model parameters and then demonstrate the behavior of the full distribution $f_{\rm comp}(t)$.

\subsubsection{Mean and variance of the completion time} 
The $m^{\rm th}$ moment of $T_{\rm comp}$ can be calculated from the Laplace transform of $f_{\rm comp}(t)$ as
\begin{gather}
\left \langle \left( T_{\rm comp} \right)^{m} \right \rangle = (-1)^{m} \lim_{\epsilon \to 0} \frac{d^{m} \wt{f}_{\rm comp}}{d \epsilon^{m}},
\label{eqn:comp-moments}
\end{gather}
which follows from \eqref{eqn:laplace-transform-definition}. We define $\mu_{\rm comp}=\left \langle T_{\rm comp} \right \rangle$ and $\textrm{CV}(T_{\rm comp})= [ \langle ( T_{\rm comp} )^{2} \rangle -\mu_{\rm comp}^{2} ] / \mu_{\rm comp}^{2}$, as the mean and coefficient of variance of the completion time.

\begin{figure}[t]
\centering
\includegraphics{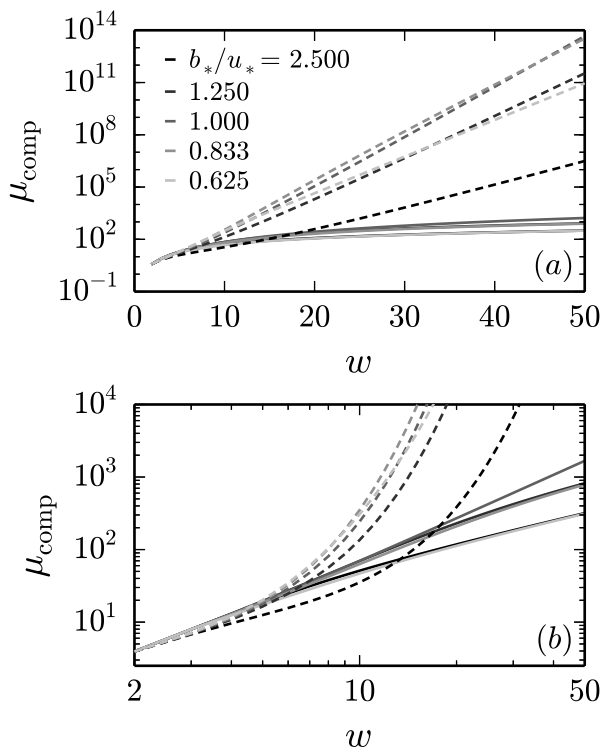}
\caption{Mean completion time as a function of the total number of molecules in the complex. In all plots, solid and dashed curves correspond to sequential and random binding, respectively. In (a), $\mu_{\rm comp}$ is plotted on semi-logarithmic axes (y), and indicates that the mean completion time asymptotically grows exponentially for random binding. In (b), $\mu_{\rm comp}$ is plotted on log-log axes, and suggests that the mean completion time asymptotically grows as a power-law, or slower, for sequential binding. Parameter values for all graphs are: $\gamma=0.1$, $b_1=2$, $b_*=0.25$, and $u_1=1$, measured in units of $f$, the hopping rate along the filament.}
\label{fig:mu-comp-w}
\end{figure}

Fig. \ref{fig:mu-comp-w} shows $\mu_{\rm comp}$ as a function of $w$, for a set of $b_{*}/u_{*}$ values. Plots in (a) and (b) correspond to identical parameter values, but the axes are scaled differently. Looking at (a), where the y-axis is scaled logarithmically, we first note that $\mu_{\rm comp}$ increases exponentially with $w$ for the case of random binding (dashed curves). The exponent increases as $b_{*}/u_{*}$ decreases to approach the value 1 from above, which is simple to grasp intuitionally, as higher relative unbinding rates would lead to longer completion times. When the ratio $b_{*}/u_{*}$ is below one, results show the opposite trend, where the exponent decreases with decreasing $b_{*}/u_{*}$. This reflects the conditional nature of the completion time. In the previous section, we showed that $P_{\rm comp}$ approaches 0 for $b_*/u_*<1$ as $w$ increases, meaning that the fraction of trajectories that lead to the completion of the process becomes negligible. Results in Fig. \ref{fig:mu-comp-w} (a) suggest that while it is quite unlikely for the process to complete when $b_{*}/u_{*}<1$, the trajectories that do lead to completion take shorter and shorter times as $u_{*}$ increases (this point will be discussed further below). Note that $\mu_{\rm comp}$ clearly increases slower than exponentially for sequential binding (solid curves).

In Fig. \ref{fig:mu-comp-w} (b), the same data is plotted on logarithmic axes. The most prominent feature here is that $\mu_{\rm comp}$ for sequential binding (solid curves) increases as a power law for $b_{*}/u_{*}=1$, and slower than a power law for $b_{*}/u_{*} \neq 1$. The rate of increase as a function of $b_{*}/u_{*}$ follows a trend akin to that for random binding in (a).

\begin{figure}[t]
\centering
\includegraphics{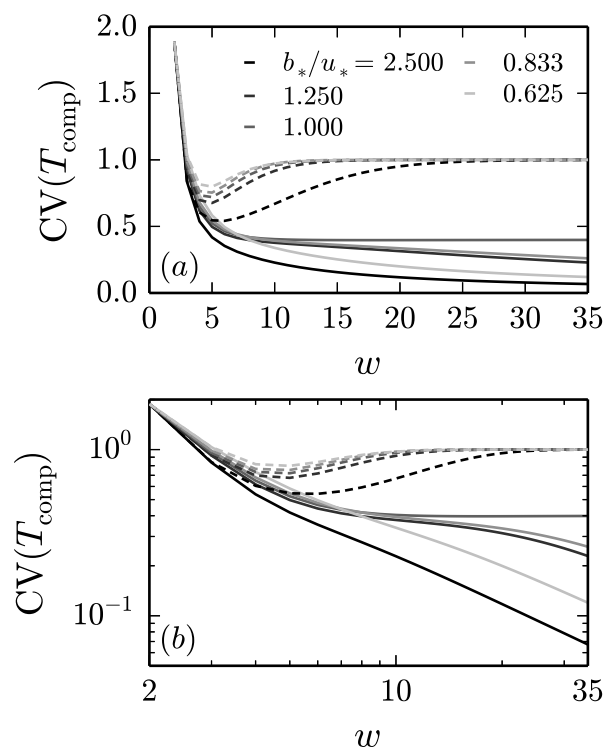}
\caption{Coefficient of variance of the completion time for sequential and random binding models. Solid and dashed lines correspond to sequential and random binding, respectively, for various values of the ratio $b_*/u_*$. Parameter values are: $\gamma=0.1$, $b_1=1$, $b_*=0.25$, and $u_1=1$, measured in units of $f$, the hopping rate along the filament.}
\label{fig:CV-vs-w}
\end{figure}

Results above show that the behavior of the average completion time is qualitatively different for random and sequential binding models as the number of molecules in the complex increases. Since completion of the assembly is a stochastic process with many steps, one may expect significant variance in the values of $T_{\rm comp}$ such that a typical value may lie far away from $\mu_{\rm comp}$. To investigate this, we calculate the coefficient of variance (CV) of $T_{\rm comp}$, defined just below \eqref{eqn:comp-moments}. Fig. \ref{fig:CV-vs-w} shows $\textrm{CV}(T_{\rm comp})$ as a function of $w$ for the same set of $b_{*}/u_{*}$ values in Fig. \ref{fig:mu-comp-w}. As seen in Fig. \ref{fig:CV-vs-w} (a), we find that random and sequential binding models display approximately the same level of variability in $T_{\rm comp}$ for the first few $w$ values. However, for $w \gg 2$, the coefficient of variance can be significantly smaller for sequential binding compared with random binding. In random binding (dashed curves), CV transiently dips below 1 for a range of $w$ values depending on $b_{*}/u_{*}$, but eventually approaches the value 1. In sequential binding however, the asymptotic behavior of CV depends on the value of $b_{*}/u_{*}$, as shown in the log-log plot in Fig. \ref{fig:CV-vs-w} (b). 


When $b_*/u_*=1$, random and sequential binding models both give rise to a CV that attains a constant value as $w$ increases, that is, the standard deviation of $T_{\rm comp}$ increases at the same rate as its mean. This implies that the distribution of completion times remain well-dispersed no matter how large the number of auxiliary molecules gets. 

In the case of sequential binding, for $b_{*}/u_{*} \neq 1$, we observe that the CV decays to zero with $w$. While it is not straightforward to provide a simple explanation for this behavior for all values of $b_{*}/u_{*}$, we can gain some insight into it by considering the extreme cases $b_*/u_*\gg 1$ and $b_*/u_*\ll 1$, as discussed below.

When $b_*/u_*\gg 1$, almost all transitions are towards completion, meaning that we have $\textrm{mean}\left(T_{\rm comp}\right) \approx wb_*^{-1}$, that is, the number of transitions to completion multiplied by the average duration of a transition. Since the variance of the duration between transitions (towards completion) is given by $b_*^{-2}$, variance of the completion time would be $\textrm{var}\left( T_{\rm comp} \right) \approx wb_*^{-2}$. Therefore, we expect the CV to decay as $w^{-1}$ in the limit $b_*/u_*\gg 1$, which is consistent with the power-law behavior observed in Fig. \ref{fig:CV-vs-w} (b). 

When $b_*/u_* \ll 1$, we expect the process to reach completion very rarely, as the complex is much more likely to shrink than grow at any state. Nevertheless, when the complex does form, the maximally likely system trajectory would consist of $w$ consecutive transitions towards completion, since any transition towards the unbound state would introduce an additional multiplicative factor of $b_*/u_*$ in the likelihood of a trajectory that reaches completion. Therefore, we expect CV to decay in the same way as it does for $b_*/u_*\gg 1$, as explained in the paragraph above; the only difference is that completion only rarely occurs for $b_*/u_* \ll 1$, while it is the typical outcome for $b_*/u_* \gg 1$.


We note that a previous study by D'Orsogna and Chou~\cite{dorsogna_first_2005} employed a similar model, although in a different context (ligand-receptor binding) and without spatial extent, and found that random binding results in faster mean completion times compared with sequential binding, except when $b_*/u_* \approx 1$, in the absence of any cooperativity. Our findings are consistent with this result up to a certain value of $w$, say $w_{\rm c}$. As seen in Fig. \ref{fig:mu-comp-w} (b), the crossover value $w_{\rm c}$ above which random binding leads to longer completion time increases with $b_*/u_*$. We also would like to point out that the presence of diffusive states and the possibility of dissociation significantly affects the behavior of $P_{\rm comp}$.

Overall, we find that sequential binding model results in more precise timings compared to the random binding model. In addition, as clearly seen in Fig. \ref{fig:mu-comp-w}, sequential binding is orders of magnitude faster than random binding for $w\gtrsim 10$. Note that the predictions of an assembly model is only feasible if the completion time is less than the longest time scale in a cell, i.e. duration of the cell cycle. 

\subsubsection{Distribution of the completion time}
Here we demonstrate the full distribution of $T_{\rm comp}$, which is obtained by taking the inverse Laplace transform of the expression in \eqref{eqn:fcomp-laplace} (see Appendices \ref{sec:PcompTcomp} and \ref{sec:inverse-laplace}). 

\begin{figure}[b]
\centering
\includegraphics{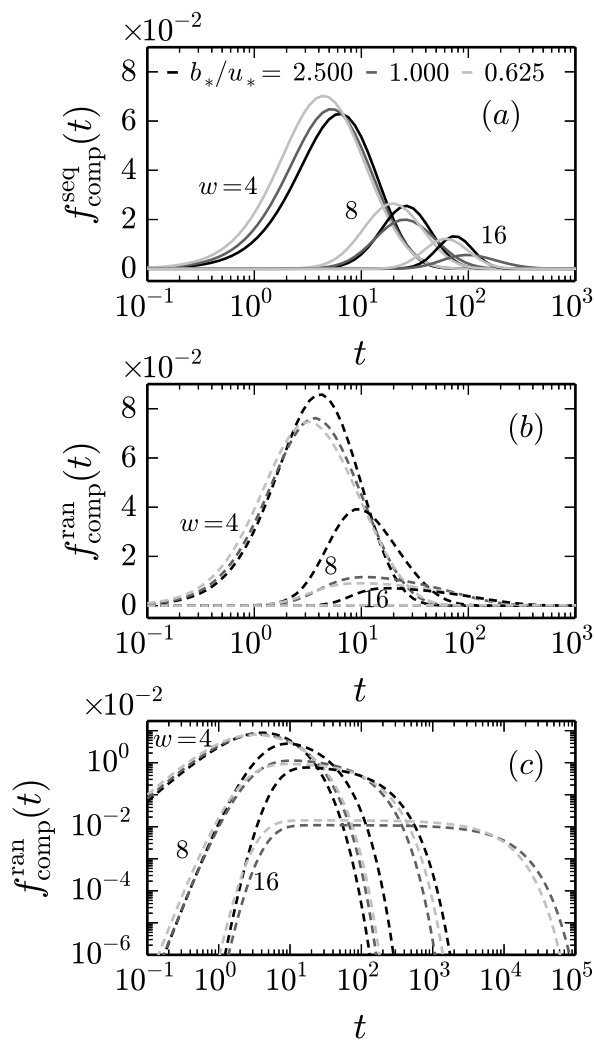}
\caption{Distribution of the completion time for different values of the total number of molecules and the ratio $b_*/u_*$. Parameter values are: $\gamma=0.1$, $b_1=2$, $b_*=0.25$, and $u_1=1$, measured in units of $f$, the hopping rate along the filament.}
\label{fig:Tcomp-dist}
\end{figure}

Fig. \ref{fig:Tcomp-dist} (a-c) show $f_{\rm comp}(t)$ for sequential and random binding models, indicated by the superscripts ``seq'' and ``ran''. All graphs display $f_{\rm comp}(t)$ as a function of $t$, for $w=4,\,8,\,16$ and $b_{*}/u_{*}=2.5,\, 1,\, 0.625$, where different $b_{*}/u_{*}$ values are color coded. Note that time axes are logarithmically scaled and cover a broad range containing 4 to 6 decades. Looking at (a), we see that $f_{\rm comp}^{\rm seq}(t)$ has a bell-shaped form, whose peak increases with $w$ as one would expect. We observe that the distribution is most widely spread around its peak when $b_{*}/u_{*}=1$. This is consistent with the CV shown in Fig. \ref{fig:CV-vs-w}, for $w \gg 2$, where CV for sequential binding attains its maximum at $b_{*}/u_{*}=1$, and decreases for other values, implying a narrower distribution. In (b), identical parameter values are used to plot $f_{\rm comp}^{\rm ran}(t)$. This time, we note that the distribution gets spread over a wide range of $t$ much quicker with increasing $w$. To better visualize this case, we display the same data on logarithmically scaled axes in (c). We immediately note that the curves for $w=16$ are now clearly visible, and that the distribution becomes significantly uniform over a broad range of $t$ values as $w$ increases (for instance, $f_{\rm comp}^{\rm ran}(t)$ for $w=16$ attains the value $\approx 10^{-4}$ over a range $\approx 10^{4}$, implying that almost all of the probability is contained in the plateau). This behavior is also in agreement with the results obtained in Fig. \ref{fig:CV-vs-w}, where CV for random binding approaches 1 with increasing $w$, indicating that the distribution remains well-spread over a range of $t$ that grows with $\mu_{\rm comp}$.

\subsection{Residence time in an interval}
\label{sec:residence-time}
Using techniques of single molecule microscopy, it is possible to directly observe trajectories of individual molecules, especially in one and two dimensions. Analyzing trajectories that exhibit binding and unbinding events, one can extract useful information about reaction kinetics at the molecular level~\cite{koyama-honda_fluorescence_2005,kasai_full_2011}. Nevertheless, this almost always requires fitting a model to the data.

If a labeling method can be developed to directly observe the completion time of a molecular complex, model predictions presented so far can be used to fit experimental data and extract kinetic parameters. Nevertheless, this would be a challenging task. More often than not, there is uncertainty in the number of molecules in the complex as well as in determining which molecule(s) would best characterize the completion of the complex.

In this section, we present model predictions for the amount of time the seed spends in an interval of length $2r$ centered around the target site, given it was initially bound and did not dissociate until the time of measurement. We refer to this time as the \emph{residence time}, which is also the first-passage time of the seed at a distance $r$ from the target site, given dissociation is prevented by setting $\gamma=0$. The choice $\gamma=0$ is justified, as a particle cannot be tracked anymore after it dissociates. Therefore, all measurements of $T_{\rm res}$ thus defined corresponds to a model where $\gamma=0$. Although the information contained in the residence time is more indirect compared with the information that would be contained in the completion time, the residence time is probably easier to measure in practice.

To calculate the distribution of the residence time, denoted by $f_{\rm res}(t)$, we consider the limit $\kappa \to \infty$ in the model illustrated in Fig. \ref{fig:illustration-Markov-chain}, which amounts to placing perfectly absorbing boundaries at $m=-r$ and $r$. Supposing that the seed molecule does not dissociate, we can calculate the first-passage time at site $-r$ or $r$, from the knowledge of $P_{m}(t)$ and $Q_{i}(t)$. Details of the calculation are presented in Appendix \ref{sec:Tres}. Formulas for the mean and variance of $f_{\rm res}(t)$ for the first few $w$ are also given in Appendix \ref{sec:Tres}.

\begin{figure}[t]
\centering
\includegraphics[scale=1]{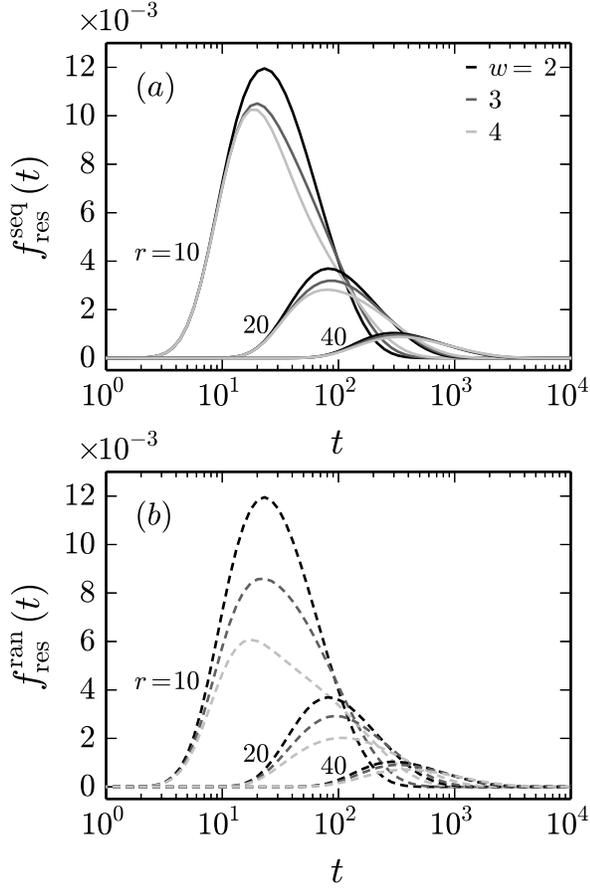}
\caption{Distribution of the residence time for sequential and random binding models. Each graph shows groups of three curves, characterized by the same $w$ value, corresponding to three different values of the interval length $2r$, measured in units of lattice spacing. (a) and (b) show $f_{\rm res}(t)$ as a function of $t$ for sequential and random binding, respectively. Parameter values are: $\gamma=0.1$, $b_1=2$, $b_*=0.25$, and $u_1=1$, measured in units of $f$, the hopping rate along the filament.}
\label{fig:residence-time}
\end{figure}

Fig. \ref{fig:residence-time} shows $f_{\rm res}(t)$ as a function of $t$ for sequential (a) and random (b) binding models, for the first few values of $w$. We observe that the profile of $f_{\rm res}(t)$ contains useful information about the binding model at smaller values of $r$. As $r$ gets larger, distributions with different parameter values start to look similar. In the presence of measurement errors, this would make it difficult to infer the molecular details about the assembly process via measurement of $T_{\rm res}$.

One possible way of directly observing the residence time could be achieved by employing nano-materials such as DNA origami frames. Two-dimensional frames made up of DNA that contain a stretched filament can be observed with atomic force microscopy as well as light microscopy~\cite{yamamoto_single_2014}. This allows making measurements in a virtually 2-D space such that the seed molecule would not go out of focus while it is bound to the filament.

\subsection{When multiple seed molecules are present}
\label{sec:multiple-seeds}
We expect the medium to contain multiple seed molecules undergoing facilitated diffusion such that the overall rate of completion depends on how frequently a \emph{new} seed molecule binds to the target site. By a \emph{new} molecule, we mean any molecule except the one that has not dissociated from the filament after becoming unbound at the target site (following an incomplete assembly). When the concentration of the seed molecules is low, competition among different molecules for the target site is approximately negligible. In this case, arrival of new molecules at the first bound state can be approximated by a Poisson process, where the time until arrival is distributed exponentially. In other words, the system attempts to complete the assembly process at constant rate. We would like to remark that this approximation was considered in similar contexts in refs.~\cite{kaizu_berg-purcell_2014} and \cite{paijmans_lower_2014}, where the low concentration assumption is also discussed in the light of biologically relevant values of molecular concentrations.  

Let $T_{\rm dis}$ denote the time at which a seed that started in the first bound state dissociates, regardless of whether the process completes or not. Distribution of $T_{\rm dis}$, denoted by $f_{\rm dis}$ is given in Appendix \ref{sec:Tdis}. Next, we define $f_{\rm arr}(t)$ and $f_{\rm dis}^{\prime}(t)$ to be distributions of the first arrival time, $T_{\rm arr}$, of a new seed at the first bound state, and the first dissociation time, $T_{\rm dis}^{\prime}$, of a seed that was initially bound and assuming that the process is not allowed to complete, that is, $b_w \to 0$. Note that the time $T_{\rm dis}^{\prime}$ is introduced for convenience, and its role in calculating the completion time will become clear in what follows.  

Suppose that initially there are no molecules in the vicinity of the target site. After a time $T_{\rm arr}$, we expect a molecule to become bound, which would lead to completion with probability $P_{\rm comp}$ after a time $T_{\rm comp}$, or to dissociation without completion with probability $1-P_{\rm comp}$, after time $T_{\rm dis}^{\prime}$. If we call this an \emph{attempt}, then we can formulate the distribution of the first completion time in terms of the number of attempts that lead to the completion of the process. Note that this requires $\left\langle T_{\rm dis}^{\prime} \right\rangle \ll \left\langle T_{\rm arr} \right\rangle$, implying that binding of a new molecule while another has not yet dissociated from the filament is unlikely. 

Let $g_{\rm comp}(t)$ be the distribution of the completion time when the assumption above holds. We can express $g_{\rm comp}(t)$ as
\begin{align}
g_{\rm comp}(t) &= P_{\rm comp} \left[ f_{\rm arr} \ast f_{\rm comp} \right] \nonumber \\
				&+ (1-P_{\rm comp})P_{\rm comp} \left[ f_{\rm arr} \ast f_{\rm dis}^{\prime} \ast f_{\rm arr} \ast f_{\rm comp} \right] \nonumber \\
				&+ (1-P_{\rm comp})(1-P_{\rm comp})P_{\rm comp} \nonumber \\ 
				&\ \ \times \left[ f_{\rm arr} \ast f_{\rm dis}^{\prime} \ast f_{\rm arr} \ast f_{\rm dis}^{\prime} \ast f_{\rm arr} \ast f_{\rm comp} \right] \nonumber \\ 
				& + \cdots, 
	\label{eqn:g-comp-formal}
\end{align} 
where $\ast$ denotes convolution, that is, $f\ast g = \int_0^{t} ds f(t-s)g(s)$. In \eqref{eqn:g-comp-formal}, the first, second and third terms correspond to the probability that the process is completed after the first, second and third attempt, multiplied by the distribution of the time each route takes. Taking the Laplace transform of \eqref{eqn:g-comp-formal}, thereby converting convolutions into products, we obtain
\begin{align}
\wt{g}_{\rm comp}(\epsilon) &= \frac{P_{\rm comp} \wt{f}_{\rm comp}(\epsilon)}{ \left(1-P_{\rm comp} \right)\wt{f}^{\prime}_{\rm dis}(\epsilon)} \nonumber \\ &\times\sum_{n=1}^{\infty} \left[ \left(1-P_{\rm comp}\right) \wt{f}_{\rm arr}(\epsilon) \wt{f}_{\rm dis}^{\prime}(\epsilon) \right]^n \nonumber \\
	&= \frac{P_{\rm comp} \wt{f}_{\rm arr}(\epsilon) \wt{f}_{\rm comp}(\epsilon) }{1 - \left( 1-P_{\rm comp}\right) \wt{f}_{\rm arr}(\epsilon) \wt{f}_{\rm dis}^{\prime}(\epsilon) },
	\label{eqn:g-comp-laplace}
\end{align}
where we assumed the convergence of the geometric sum, which certainly holds as $\epsilon \to 0$, since $0 \le P_{\rm comp} \le 1$, and $\wt{f}_{\rm arr}(\epsilon)$ and $\wt{f}_{\rm dis}^{\prime}(\epsilon)$ are Laplace transforms of normalized probability distributions.

Since we assume that the arrival of a new seed is a Poisson process, we have $f_{\rm arr}(t)=\alpha e^{-\alpha t}$, where $\alpha$ is a function of the 3-D diffusion coefficient as well as the unspecific binding/unbinding rates between the seed and the filaments. The distribution $f_{\rm dis}^{\prime}(t)$ can be calculated as described in Appendix \ref{sec:Tdis}.

We can then calculate the mean completion time, starting from the state where no seed is bound, as
\begin{gather*}
\left\langle T_{\rm comp} \right\rangle = -\lim_{\epsilon \to 0} \frac{d \wt{g}_{\rm comp}}{d\epsilon},
\end{gather*}
which follows from \eqref{eqn:comp-moments}. Substituting \eqref{eqn:g-comp-laplace} into the equation above and performing the limit, we get
\begin{gather*}
\left\langle T_{\rm comp} \right\rangle = \mu_{\rm arr} + \mu_{\rm comp} + \left( \frac{1-P_{\rm comp}}{P_{\rm comp}} \right) \left( \mu_{\rm arr} + \mu_{\rm dis^{\prime}} \right),
\end{gather*}
where $\mu_{\rm arr}=\langle T_{\rm arr} \rangle = \alpha^{-1}$ and $\mu_{\rm dis^{\prime}} = \langle T_{\rm dis^{\prime}} \rangle$, where the angular brackets imply expected values, as in \eqref{eqn:comp-moments}. If the complex disappears from the system, then we can estimate the rate of formation as
\begin{gather*}
k_{\rm comp} = \frac{1}{\left\langle T_{\rm comp} \right\rangle}.
\end{gather*}

We note that the theory developed by Berg et al.~\cite{berg_diffusion-driven_1981}, or possibly an extension of the widely applicable mean first-passage time calculations given by B\'{e}nichou et al.~\cite{benichouo._geometry-controlled_2010}, can be used to approximate the association rate $\alpha$ as a function of parameters characterizing the whole system, including the 3-D diffusion coefficient and the ratio of the average inter-filament distance to filament radius.

\section{Discussion and Conclusions}
\label{sec:discussion}
In this work, we presented a model and its analytic solution for the formation of a complex of molecules on a filament, applicable to the assembly of the transcription complex. The process is driven by seed molecules that undergo facilitated diffusion, which consists of 3-D diffusion interrupted by episodes where a molecule associates with a filament and undergoes 1-D diffusion in search of a target site. In this work, we did not explicitly model the 3-D diffusion of a seed molecule, which was studied earlier~\cite{berg_diffusion-driven_1981}. Once the seed molecule becomes bound to the target site, a number of auxiliary molecules can reversibly bind, forcing the seed to stay bound, until an assembly of a given size forms. We believe that including spatial degrees of freedom and accounting for the effect of facilitated diffusion is the major contribution of this study to the existing body of work on the kinetics of aggregation.

Two quantities were used to characterize the process: 1) the probability that the assembly completely forms and 2) the time it takes for the process to complete, conditioned on completion; once a seed molecule becomes bound. Mathematical expressions for these quantities are given in \eqref{eqn:Pcomp} and \eqref{eqn:fcomp-laplace}, respectively. In Appendix \ref{sec:simulation}, we also provide a comparison of analytical expressions with simulations, verifying the validity of the results.

Similar to previous studies (see, for instance, \cite{dorsogna_first_2005} and \cite{luijsterburg_stochastic_2010}, where the latter has an experimental component), we found that the order in which auxiliary molecules bind would matter, and compared two different models where auxiliary molecules bind in a strictly sequential order and in completely random order.

The findings indicate that the probability of completion is greater for random binding than it is for sequential binding (see Figs. \ref{fig:Pcomp-vs-w} and \ref{fig:Pcomp-vs-u}). Interestingly, in random binding, when the unbinding rates of auxiliary molecules are relatively larger than their binding rates, there can be an optimal size for the complex for which the chance of completion is maximal (see Fig. \ref{fig:Pcomp-vs-w}). While the probability of completion is larger for the random binding model, it can take a long time to reach completion, especially when the complex contains much more than a few molecules. Calculating the completion time distribution for the two models, we found that the mean completion time grows exponentially with the size of the complex for the random binding model, while it grows as a power-law or slower for the sequential binding model (see Fig. \ref{fig:mu-comp-w}). In addition, completion time is much more broadly distributed for random binding compared to that in sequential binding (see Figs. \ref{fig:CV-vs-w} and \ref{fig:Tcomp-dist}). 

Therefore, there is a trade-off between the probability of completion and the completion time, and an optimal strategy may consist of an hybrid model, where the first few molecules bind in random order to stabilize the forming complex, and the rest of the molecules bind sequentially to reduce the completion time, ensuring that it does not scale exponentially with the number of molecules in the complex. 

Facilitated diffusion enhances the probability of completion by increasing the chances for the seed molecule to quickly rebind to the target site even if it becomes unbound before the process completes (this so-called \emph{rebinding} effect was also discussed by other authors, for instance, in the context of enzymatic reactions~\cite{takahashi_spatio-temporal_2010}, and in gene expression~\cite{meyer_geometry-induced_2012}). We quantified the amount of enhancement by deriving a formula for it as a function of all model parameters [see \eqref{eqn:fast-rebinding-efficiency}]. Inside the parameter range considered here, facilitated diffusion is found to enhance the completion probability more strongly for the sequential binding model compared with the random binding model (see Fig. \ref{fig:rho-contour}).

Note that the current model can be generalized to also let auxiliary molecules undergo facilitated diffusion. One possible way of achieving this is generalizing the previously studied \emph{island growth model}~\cite{brilliantov_nonscaling_1991, kallabis_island_1998}, where monomers adsorb to a surface and undergo diffusion limited aggregation to form immobile islands, to allow for dissociation and re-adsorption of monomers.

Our results are relevant for the case where the concentration of seed molecules is sufficiently low so that the competition for the target site is negligible (also see refs.~\cite{kaizu_berg-purcell_2014} and \cite{paijmans_lower_2014}). Under this assumption, we also provided an approximate analytic expression for the Laplace transform of the completion time distribution for multiple seed molecules, from which we obtained the average completion time.

While the model we consider here is appealing for artificial systems where, for instance, stretched out DNA filaments are placed in nano-engineered structures~\cite{endo_direct_2012,yamamoto_single_2014}, applicability of the model to the formation of the transcription complex in vivo depends on the validity of two key assumptions. First, the DNA is assumed to be a 1-D filament in the vicinity of a target site (a regulatory sequence of a gene). If the dissociation rate $\gamma$ is much larger than the hop rate $f$, this assumption is more likely to hold, since the seed would not be able to explore a large section of the filament at a time. We should note that DNA can be packed inside cells in a highly organized manner; it could be condensed by proteins in bacteria, and is organized in chromosomes in eukaryotic cells~\cite{alberts_molecular_2007}. In eukaryotic cells, before transcription of a gene begins, the structure of the DNA around the regulatory region of the gene loosens up such that transcription factors can directly bind to the base pairs. How well this loosened up section of the chromatin can be approximated by a filament should eventually be verified by experiments in vivo. Second, we assume that the hopping rate along the DNA is uniform. Nevertheless, recent studies showed that DNA-binding molecules can act like ``road blocks'' that can hinder the 1-D diffusion of transcription factors~\cite{li_effects_2009}. In the presence of molecules that act as road blocks as well as non-uniformity in the rate of diffusion along different sections of the DNA due to other reasons, the model considered here may underestimate, for instance, the variance and higher moments of the completion time distribution. To improve on this point, one can use a more involved model of diffusion in 1-D, allowing for random transition rates, and permeable barriers~\cite{powles_exact_1992,dudko_diffusion_2004,kalay_effects_2008,kalay_effects_2012}, and results derived for transport in random environments~\cite{kenkre_reineker_1982,hughes_random_1996}.

In this work, we did not investigate the presence of cooperativity in binding rates and assumed that the motion of molecules can be described by Brownian diffusion (continuous time random walk with exponential waiting times). Inclusion of cooperative binding/unbinding of auxiliary molecules and considering anomalous diffusion can affect the stability of the complex being formed~\cite{dorsogna_first_2005}, and lead to correlated bursts in gene expression~\cite{meyer_geometry-induced_2012}.

A key factor that determines the efficiency of search for target sites via facilitated diffusion is the relative affinity of the seed to unspecific sites on the DNA compared with that to the target site. It is worthwhile to note that this point becomes even more significant when one accounts for the degradation of DNA-binding molecules~\cite{burger_abduction_2010}. In our model, the seed hops between all adjacent sites, including the target site, at the rate $f$, and in all of the plots we choose $b_1=2f$, implying that binding to the target site happens at only twice the rate at which the seed hops between any two sites. Therefore, results illustrated here correspond to the case where the target site is not strongly ``sticky''. Note that this is consistent with experimental findings showing that the probability of binding at the first encounter is not necessarily close to 1~\cite{hammar_lac_2012}.

In Section \ref{sec:residence-time} we presented the model prediction for the time it takes for a seed to escape from an interval around the target site, provided that it does not dissociate from the filament until it is observed to escape. Single molecule observations can be performed by tagging multiple molecules and can be sophisticated enough to provide direct information about molecular interactions (for a review on multiple experimental methods, see, for instance, ref.~\cite{mueller_quantifying_2013}). Nevertheless, in the most basic setting, assuming that the seed molecule can be tracked, the dissociation time and the residence time for the seed can be directly accessed, providing evidence for performing model selection. However, we remark that the applicability of an analysis using the residence time defined above would often be limited to observations in the vicinity of the binding site, and may require high position precision. This is because the in vivo unspecific binding strength of molecules to the DNA filament is not necessarily high enough for the molecule to cover a large distance on the DNA without being dissociated. To provide a quick estimate, we expect the diffusive displacement of a molecule to go as $d=\sqrt{ \langle m^2 \rangle} \sim \sqrt{f \tau}$, where $\tau$ is the typical time a molecule spends attached, which can be estimated as $\tau \sim 1/\gamma$. Therefore, the distance a molecule covers typically goes as $d \sim \sqrt{f/\gamma}$. Nevertheless, several single molecule observations of RNA polymerase in vitro suggest that the dissociation times can be long to allow significant motion of the particle along the DNA (see, for instance, refs. \cite{kim_single-molecule_2007} and \cite{endo_direct_2012}). Finally, as also noted in Section \ref{sec:residence-time}, we remark that the discriminative power of this analysis diminishes as the length of the interval increases (see Fig. \ref{fig:residence-time}).

Recent experimental studies on the kinetics of formation of the transcription initiation complex aim to test competing hypotheses of sequential and random binding. In this respect, we believe that model predictions, in conjunction with cutting edge experimental methods, would be useful for revealing the dynamics of such mesoscopic systems.

\begin{acknowledgments}
This research was supported in part by JSPS Grant-in-Aid for Young Scientists (KAKENHI Grant Number 26730150) and the World Premier International Research Center (WPI) Initiative of the Ministry of Education, Culture, Sports, Science and Technology (MEXT) of Japan.
\end{acknowledgments}

\appendix

\begin{widetext}

\section{Master equation and its solution}
\label{sec:master-eqn-solution}

In this section we present the mathematical formulation of the model illustrated in Fig. \ref{fig:illustration-Markov-chain} in terms of a set of master equations, and its solution. 

We model the seed molecule as a random walker hopping between nearest neighboring sites in a 1-D lattice (see Fig. \ref{fig:illustration-Markov-chain}). When the random walker occupies the site $m=0$, it can transition to a bound state at rate $b_1$. We consider a total number of $w$ bound states, each of which is accessible in a sequential manner, e.g. a stack of bound states. Among the bound states, the rate of transition from state $i$ to $i+1$ is denoted by $b_{i+1}$, and the rate of transition from $i$ to $i-1$ is denoted by $u_i$. When the random walker occupies the first bound state, it becomes unbound at rate $u_1$ and returns to the lattice. Note that $b_i$ and $u_i$ characterize the rate of going deeper and shallower in the stack of bound states, respectively.

Let $P_m(t)$ denote the probability of finding the random walker unbound at site $m$ and $Q_i(t)$ be the probability that it is in the $i^{\rm th}$ bound state. The master equations that govern these probabilities are given by
\begin{align}
\frac{dP_m}{dt} = f\left( P_{m-1} + P_{m+1} - 2P_m \right) -\gamma P_m + \Big[ \delta_{m,0}\left( -b_1 P_0 + u_1 Q_1 \right) - \kappa\left( \delta_{m,-r} + \delta_{m,r} \right)P_m \Big],
\label{eqn:master-eqn-0}
\end{align}
where $\kappa$ is a parameter that adjusts the strength of an absorbing boundary at $-r$ and $r$, and
\begin{align}
\frac{dQ_1}{dt} &= b_1P_0 -\left(u_1+b_2\right)Q_1 + u_2 Q_2, \nonumber \\
\frac{dQ_2}{dt} &= b_2Q_1 -\left(u_2+b_3\right)Q_2 + u_3 Q_3, \nonumber \\
&\vdots \label{eqn:dQ_dt} \\
\frac{dQ_{w-1}}{dt} &= b_{w-1}Q_{w-2} -\left(u_{w-1}+b_{w}\right)Q_{w-1} + u_{w} Q_{w}, \nonumber \\
\frac{dQ_{w}}{dt} &= b_{w}Q_{w-1} -u_{w} Q_{w}. \nonumber 
\end{align}
We start with the solution for $P_m(t)$. Let us denote by $\varphi_{m-n}(t)$, the solution of \eqref{eqn:master-eqn-0} with the initial condition $P_m(0)=\delta_{m,n}$, when the terms inside the square brackets are set to zero, which corresponds to random walk in an infinite lattice where the random walker disappears at rate $\gamma$. Using Laplace and discrete Fourier transforms to convert differential equations to algebraic equations, $\varphi_{m-n}(t)$ can be obtained as
\begin{gather}
\varphi_{m-n}(t) = e^{-\left( 2f + \gamma \right)t} I_{m-n}(2ft),
\label{eqn:diffusion-propagator}
\end{gather}
where $I_m(t)$ denotes the modified Bessel function of the first kind~\cite{abramowitz_handbook_1964}. The Laplace transform of \eqref{eqn:diffusion-propagator} is given by (see ref.~\cite{roberts_table_1966} pp. 75)
\begin{gather}
\mathcal{L} \left\{ e^{-a t} I_{\nu}(b t) \right\} = \frac{\left[ (\epsilon+a+b)^{1/2} - (\epsilon+a-b)^{1/2} \right]^{2\nu} }{\left(2b\right)^{\nu} \left[ \left( \epsilon+a \right)^2 - b^2 \right]^{1/2}},
\end{gather}
where, ${\rm Re}[\nu] > -1$, ${\rm Re}[\epsilon] > \max \left\{ {\rm Re}[b-a],\, -{\rm Re}[b+a]  \right\}$ and $\mathcal{L}$ denotes the Laplace transform defined in \eqref{eqn:laplace-transform-definition} with $\epsilon$ as the Laplace variable. We use tildes $(\,\widetilde{}\,)$ to denote Laplace transformed variables as in the main text.
To obtain the full solution of \eqref{eqn:master-eqn-0}, we note that it is a first order linear differential equation, which allows us to express its solution as 
\begin{gather}
P_m = \sum_{n}P_{n}(0)\varphi_{m-n} + \int_0^{t} ds \sum_n \varphi_{m-n}(t-s)\left[ \cdots \right](n,s),
\label{eqn:P_m-formal-0}
\end{gather}
where $\left[ \cdots \right](n,s)$ corresponds to the expression in square brackets on the right hand of \eqref{eqn:master-eqn-0}, as a function of $n$ and $s$. Note that we need to express $Q_1$ in terms of $P_0$ in order to obtain a closed equation for $P_m$'s. To achieve this, we formally solve the system of equations given in \eqref{eqn:dQ_dt} for an initial condition where the first bound state is occupied with probability $Q_1(0)$ and all other bound states are initially unoccupied, that is, $Q_i(0)=0$ for $i>1$. Taking the Laplace transform of the system in \eqref{eqn:dQ_dt}, and solving recursively, we find
\begin{align*}
\wt{Q}_{w} &= \wt{Q}_{w-1} \frac{b_{w}}{\epsilon + u_{w}}, \\
\wt{Q}_{w-1} &= \wt{Q}_{w-2} \cfrac{b_{w-1}}{ \epsilon +u_{w-1}+b_{w} -\cfrac{u_{w} b_{w}}{\epsilon +u_{w}}}, \\
\wt{Q}_{w-2} &= \wt{Q}_{w-3} \cfrac{b_{w-2}}{\epsilon +u_{w-2}+b_{w-1} -\cfrac{u_{w-1} b_{w-1}}{\epsilon +u_{w-1}+b_{w}-\cfrac{u_{w} b_{w}}{\epsilon + u_{w}}}}, \\
&\,\,\,\vdots\,.
\end{align*}
Note that we readily have
\begin{gather}
\wt{Q}_2 = \wt{Q}_1 \frac{b_2}{\epsilon +u_2},
\end{gather}
which can be substituted in the Laplace transform of the first equation in \eqref{eqn:dQ_dt} to obtain an equation that only involves $\wt{Q}_1$ and $\wt{P}_0$, whose solution is
\begin{gather}
\widetilde{Q}_1 = \frac{Q_1(0) + b_1 \widetilde{P}_0}{\epsilon + u_1 + b_2 + K_2(w) u_2},
\end{gather}
where $K_j(w)$ is defined as
\begin{align}
K_i(w) &= \frac{1}{u_i} \K_{j=i}^w \frac{(-u_j b_j)}{(\epsilon+u_j+b_{j+1})} \label{eqn:K-i} \\
	&= -\cfrac{b_i}{\epsilon + u_{i}+b_{i+1} - \cfrac{u_{i+1} b_{i+1}}{\epsilon + u_{i+1}+b_{i+2} - \cfrac{u_{i+2} b_{i+2}}{\cdots -
          \cfrac{\ddots}{\epsilon + u_{w-1}+b_{w} - \cfrac{u_{w} b_{w}}{\epsilon + u_{w}}}}}} , \nonumber
\end{align}
where the big-K notation is one of the convenient ways of denoting parts of continued fractions, defined as~\cite{wiki_kettenbruch}
\begin{gather*}
\K_i^j \frac{a_i}{b_i} = \cfrac{a_i}{b_i + \cfrac{a_{i+1}}{b_{i+1} + \cfrac{a_{i+2}}{\cdots + \cfrac{\ddots}{b_{j-1} + \cfrac{a_{j}}{b_j}}}}}.
\end{gather*}
Note that in the limit $\epsilon \to 0$, we have
\begin{gather}
\lim_{\epsilon \to 0} K_i(w) = -\frac{b_i}{u_i}.
\end{gather}
Combining all these, we observe that the solution for $\wt{Q}_i$ can be compactly written as 
\begin{gather}
\widetilde{Q}_i = (-1)^{i-1}\widetilde{Q}_1 \prod_{j=2}^i K_j(w),
\label{eqn:Q_i-recursive}
\end{gather}
for $i\ge 2$. For $w\ge 2$, the total probability of being bound is given by
\begin{gather}
\sum_{i=1}^w \widetilde{Q}_i = 
\left[1 - K_2\left(1 - K_3\left(\cdots -K_{w-1}\left( 1 - K_{w} \right) \cdots \right) \right) \right] \widetilde{Q}_1,
\label{eqn:Q-sum}
\end{gather}
which is obtained by summing up (\ref{eqn:Q_i-recursive}) and rearranging terms in the summation (the argument $w$ of $K_i(w)$ is omitted for brevity in notation). When there is only a single bound state, that is, $w=1$, the probability of being bound in simply equal to $Q_1$.

Taking the Laplace transform of \eqref{eqn:P_m-formal-0}, we can obtain an algebraic equation for $P_m$ in terms of the probabilities $P_{-r}$, $P_0$ and $P_{r}$, which reads
\begin{gather}
\wt{P}_m = \Sigma^{\prime}_m - b^{\prime}_m - \kappa\wt{\varphi}_{m+r}\wt{P}_{-r} -\kappa \wt{\varphi}_{m-r}\wt{P}_{r},
\label{eqn:P-m-formal-laplace}
\end{gather}
where we defined
\begin{align*}
\Sigma^{\prime}_m &= \Sigma_m + \frac{u_1 Q_1(0)}{\epsilon + u_1 + b_2 + K_2(w) u_2 } \wt{\varphi}_m , \\
\Sigma_m &= \sum_{n}P_{n}(0)\wt{\varphi}_{m-n}, \\
b^{\prime}_m &= \left( 1 - \frac{u_1}{\epsilon + u_1 + b_2 + K_{2}(w) u_2 } \right) b_1 \wt{\varphi}_m.
\end{align*}
Substituting $m=-r,\,0,$ and $r$ in \eqref{eqn:P_m-formal-0}, we then obtain the following system of linear equations
\begin{gather}
\left[
\begin{array}{ccc}
1+\kappa\wt{\varphi}_0 & b^{\prime}_{-r} & \kappa\wt{\varphi}_{-2r} \\
\kappa\wt{\varphi}_{r} & 1+ b^{\prime}_0 & \kappa\wt{\varphi}_{-r} \\
\kappa\wt{\varphi}_{2r} & b^{\prime}_r & 1+\kappa\wt{\varphi}_0 
\end{array}
\right] 
\left[
\begin{array}{l}
\wt{P}_{-r} \\
\wt{P}_0 \\
\wt{P}_r 
\end{array}
\right] =
\left[
\begin{array}{l}
\Sigma^{\prime}_{-r} \\
\Sigma^{\prime}_0 \\
\Sigma^{\prime}_r 
\end{array}
\right],
\label{eqn:defect-system}
\end{gather}
which can be solved for $\wt{P}_{-r},\,\wt{P}_{0},$ and $\wt{P}_{r}$ to complete the whole solution.

\section{Probability of completion, first completion and dissociation time distributions}
In this section, we outline the calculation of the probability of completion before dissociation from the filament, the first completion time given completion precedes dissociation, and the dissociation time, all for a random walker that is initially occupying the first bound state. The appropriate limit in \eqref{eqn:master-eqn-0} is $\kappa\to 0$. The full solution for $\widetilde{P}_{m}$ is obtained by solving the system of equations in \eqref{eqn:defect-system} and substituting the resulting expressions in \eqref{eqn:P-m-formal-laplace}. The explicit solution is given by
\begin{align}
P_{m}(\epsilon) &= \Sigma _m +\frac{\tilde{\varphi }_m \left[u_1
   Q_1(0) -b_1 \Sigma _0
   \left(\epsilon +b_2+K_2 u_2\right) \right]}{u_1+
   \left( 1 + b_1\widetilde{\varphi }_0\right) \left(\epsilon +b_2+K_2(w) u_2\right) }.
\label{eqn:P-k-zero}
\end{align}
When there is only a single bound state, \eqref{eqn:P-k-zero} reduces to
\begin{gather*}
P_{m}(\epsilon)  = \Sigma _m+\frac{\tilde{\varphi }_m \left[u_1 Q_1(0) -\epsilon  b_1
   \Sigma _0 \right]}{u_1 + \epsilon \left( 1 + b_1 \tilde{\varphi }_0 \right)}.
\end{gather*}
Note that the expressions above hold for all initial conditions, described by $\Sigma_{m}$ and $Q_{1}(0)$. We consider the case where the random walker is initially occupying the first bound state, that is, $\Sigma_m=0$ and $Q_1(0)=1$. 

\subsection{Dissociation time distribution}
\label{sec:Tdis}
We denote the distribution of the dissociation time by $f_{\rm dis}(t)$. The molecule is allowed to visit any state any number of times; therefore, we do not have the restriction $u_{w}=0$ that we used in the calculation for the completion time. In this respect, $f_{\rm dis}(t)$ is not a conditional distribution, unlike $f_{\rm comp}(t)$. The distribution of the first dissociation time can formally be written as
\begin{gather}
f_{\rm dis}(t) = -\frac{d}{dt} \left( \sum_m P_m(t) + \sum_i Q_i(t) \right),
\label{eqn:f-dis-survival}
\end{gather}
where the quantity inside the parentheses is the survival probability, the probability that the random walker is still diffusing in the lattice, or the system is in any of the bound states. In the Laplace domain, this expression becomes
\begin{gather}
\widetilde{f}_{\rm dis}(\epsilon) = 1 - \epsilon \left( \sum_m \widetilde{P}_m  + \sum_i \widetilde{Q}_i \right). 
\label{eqn:f-dis-laplace}
\end{gather} 
The probability of being at any lattice site can be obtained by substituting $\wt{\varphi}_{m}(\epsilon)$ and summing over $m$
\begin{gather}
\sum_{m} \wt{P}_{m}(\epsilon) = \frac{u_{1}\left( \alpha + \gamma + 4f + \epsilon \right)}{\left( \alpha + \gamma + \epsilon \right)\left[ \left( \alpha + b_{1} \right) \left( \epsilon + b_{2} + u_{2} K_{2}(w) \right)  + u_{1} \alpha \right] }.
\end{gather}
Combining this with \eqref{eqn:Q-sum}, the explicit form of the dissociation time is obtained as
\begin{gather}
\widetilde{f}_{\rm dis}(\epsilon) = 1-\frac{\epsilon  \left(u_1 (4 f+\gamma +\epsilon +\alpha
   (\epsilon ))+\Lambda_{w}  \left((\gamma +\epsilon )
   b_1+\left(\gamma +\epsilon +b_1\right) \alpha (\epsilon
   )+\alpha (\epsilon )^2\right)\right)}{(\gamma +\epsilon
   +\alpha (\epsilon )) \left(b_1 \left(\epsilon
   +b_2\right)+\left(\epsilon +b_2+u_1\right) \alpha (\epsilon
   )+K_{2}(w) u_2 \left(b_1+\alpha (\epsilon
   )\right)\right)},
\label{eqn:f-dis-laplace-explicit}
\end{gather}
where $\Lambda_{w}$ is equal to the quantity in the square brackets on the right hand side of \eqref{eqn:Q-sum}.

In order to obtain the distribution of dissociation time given the process never reaches completion before dissociation, denoted by $f_{\rm dis}^{\prime}(t)$ in Section \ref{sec:multiple-seeds}), one can eliminate the constant $b_w$ and the variable $Q_w$ in the set of equations \eqref{eqn:dQ_dt} and drop the last equation, thereby removing the possibility of completion. Then, the solution of this modified master equation for the dissociation time, performed in the same way as above, would provide $f_{\rm dis}^{\prime}(t)$.


\subsection{Probability of completion and the completion time distribution}
\label{sec:PcompTcomp}
When the system transitions into the final bound state $w$, we call the process complete. We can calculate the probability of completion and the first completion time by setting $u_w$, the rate of leaving the final bound state, to zero and finding out the probability $Q_w$. Note that this is equivalent to setting $u_w=0$ in the master equation and solving for $Q_w$. With this choice, system trajectories that arrive at the final bound state cannot leave, and the distribution of first completion times can be written as
\begin{align}
f_{\rm comp}(t) &= \frac{1}{P_{\rm comp}} \left\{ -\frac{d}{dt}\left[ 1 - Q_{w} (t;\, u_w=0) \right] \right\}, \label{eqn:f-comp-formal-0} \\
& =\frac{1}{P_{\rm comp}} \left. \frac{d Q_{w}}{dt} \right|_{u_w=0},
\label{eqn:f-comp-formal}
\end{align}
where $P_{\rm comp}$ is the probability of completion, acting as a normalization constant. Note that the quantity in square brackets in \eqref{eqn:f-comp-formal-0} is the survival probability, defined as the probability of not having visited the $w^{\rm th}$ bound state up to time $t$. In this respect, \eqref{eqn:f-comp-formal-0} is analogous to \eqref{eqn:f-dis-survival}, except that $f_{\rm comp}$ is a conditional distribution and needs to be normalized with the probability of completion, $P_{\rm comp}$. In Laplace domain, we have
\begin{gather}
\widetilde{f}_{\rm comp}(\epsilon) = \frac{\epsilon \widetilde{Q}_{w}(\epsilon;\,u_w=0)}{P_{\rm comp}} ,
\label{eqn:f-comp-formal-laplace}
\end{gather}
since $Q_{w}(0)=0$. Note that this is equivalent to solving the master equations \eqref{eqn:master-eqn-0} and \eqref{eqn:dQ_dt} by setting $u_w=0$ first, and then calculating $Q_w$, since the solutions \eqref{eqn:Q_i-recursive} and \eqref{eqn:P-k-zero} are valid for $u_w=0$.

$P_{\rm comp}$ is the probability of occupancy of state $w$ in the long time limit. Therefore, we have $P_{\rm comp} = Q_{w}(t \to \infty;\, u_w=0)$. To find $P_{\rm comp}$ we first calculate $\wt{Q}_w(\epsilon;\, u_w=0)$ from \eqref{eqn:Q_i-recursive}. Then, using the limit theorem for the Laplace transform, $\lim_{t\to\infty} f(t) = \lim_{\epsilon \to 0} \epsilon \wt{f}(\epsilon)$, we obtain the explicit result
\begin{gather}
P_{\rm comp} = \frac{ 1 }{ 1 + \lambda(w) / \left(1 + \beta \right) },
\end{gather}
where
\begin{align}
\beta &= b_1 / \sqrt{\gamma(\gamma + 4f)}, \\
\lambda(w) &= \left[ \pi_{b}(2;w) \right]^{-1} u_1( \pi_{b}(3;w) + u_2( \pi_{b}(4;w) + \cdots + u_{w-2}( \pi_{b}(w;w) + u_{w-1} ) \cdots ) ), \label{eqn:lambda-w}\\
\pi_{b}(i;w) &= \prod_{k=i}^{w}b_k.
\end{align}
Note that $w\geq2$ and $\lambda(2) = u_1$. 

The Laplace transform of the first completion time valid for $w\geq 2$ is obtained from \eqref{eqn:f-comp-formal-laplace}, and is explicitly given by
\begin{align}
\widetilde{f}_{\rm comp}(\epsilon) = \left. \frac{(-1)^{w-1}}{P_{\text{comp}}} \left( \prod _{i=2}^w K_i(w)  \right) \left( \frac{\epsilon}{\epsilon + b_2 + u_1 \alpha(\epsilon)/\left( \alpha(\epsilon)+b_1 \right) + u_2 K_2(w)} \right)  \right|_{u_w = 0},
\label{eqn:f-comp-laplace-appendix}
\end{align}
where $\alpha(\epsilon) = \sqrt{(\gamma +\epsilon ) (\gamma +4 f+\epsilon )}$, and $K_{i}(w)$ is defined in \eqref{eqn:K-i}. 

\section{Residence time distribution}
\label{sec:Tres}
In this section, we are concerned with the case where the random walker is initially occupying the first bound state and can be absorbed if it travels far enough from the binding site. We consider the residence time of the random walker in a symmetric interval centered around the target site, given the walker does not dissociate from the lattice before exiting this interval. To calculate the residence time (equivalently, the first exit or escape time), we consider \eqref{eqn:master-eqn-0} in the limit $\kappa \to \infty$ and $\gamma \to 0$, which amounts to placing perfectly absorbing boundaries at $m=-r$ and $r$, and eliminating the possibility of dissociation from the lattice. 

After solving for $\wt{P}_m$ using the same method presented in the previous sections, the Laplace transform of the residence time distribution, which we denote by $\wt{f}_{\rm res}$, is then found by using \eqref{eqn:f-dis-laplace}. The probability of finding the molecule in the unbound state is obtained by summing $\widetilde{P}_{m}$ over all lattice sites, resulting in
\begin{align*}
\sum_{m}P_{m} &= \frac{u_1 (4 f+\epsilon +\delta (\epsilon )) \left(-(2
   f)^r+(2 f+\epsilon +\delta (\epsilon
   ))^r\right)^2}{(\epsilon +\delta (\epsilon )) \left(u_1
   \left((2 f)^{2 r}+\beta (\epsilon )\right) \delta
   (\epsilon )-\left(\epsilon +b_2+
   u_2 K_2(w) \right) \left((2 f)^{2 r} \left(b_1-\delta (\epsilon
   )\right)-\beta (\epsilon ) \left(b_1+\delta (\epsilon
   )\right)\right)\right)}
\end{align*}
where $\beta(\epsilon)=\left(\sqrt{\epsilon  (4 f+\epsilon )}+2 f+\epsilon \right)^{2 r}$ and $\delta(\epsilon)=\sqrt{\epsilon  (4f+\epsilon )}$. The probability of being in any bound state is given by the sum in \eqref{eqn:Q-sum}. Substituting these two sums in \eqref{eqn:f-dis-laplace} and performing the algebra, we obtain 
\begin{align}
\widetilde{f}_{\rm res}(\epsilon) =  1 
   &-\frac{\epsilon  \left(u_1 (4 f+\epsilon +\delta (\epsilon
   )) \left((2 f)^r-(2 f+\epsilon +\delta (\epsilon
   ))^r\right)^2 + \Lambda_w (\epsilon +\delta (\epsilon )) \left((2
   f)^{2 r} \left(-b_1+\delta (\epsilon )\right)+\beta
   (\epsilon ) \left(b_1+\delta (\epsilon )\right)\right)
   \right)}{(\epsilon +\delta (\epsilon
   )) \left(u_1 \left((2 f)^{2 r}+\beta (\epsilon )\right)
   \delta (\epsilon )+\left(\epsilon
   +b_2+ u_2 K_2(w) \right) \left((2 f)^{2 r}
   \left(-b_1+\delta (\epsilon )\right)+\beta (\epsilon )
   \left(b_1+\delta (\epsilon )\right)\right)\right)},
\label{eqn:f-res-laplace}
\end{align}
where $\Lambda_{w}$ is equal to the quantity in the square brackets on the right hand side of \eqref{eqn:Q-sum}. We denote the residency time as $T_{\rm res}$. The moments of $T_{\rm res}$ can be calculated from
\begin{gather}
\langle T_{\rm res}^n \rangle = (-1)^n \lim_{\epsilon \to 0}  \frac{d^n \widetilde{f}_{\rm res}}{d \epsilon^n}.
\end{gather}

The mean and variance of $T_{\rm res}$, normalized by $f^{-1}$ and $f^{-2}$ respectively ($f$ being the hopping rate along the lattice, proportional to the 1-D diffusion coefficient along the DNA), for the first few values of $w$ are given by
\begin{align*}
w=1:&  \\
	& \mu_{\rm res} = \frac{1}{2} \left( r^2 + \frac{2 + r b_1}{u_1} \right), \\
	& \sigma^2_{\rm res} = \frac{3 \left(2+r b_1\right)^2+2 \left(r+2 r^3\right) b_1 u_1+r^2 \
\left(1+2 r^2\right) u_1^2}{12 u_1^2}, \\
w=2:&  \\
	& \mu_{\rm res} = \frac{1}{2} \left[ r^2+\frac{\left(2+r b_1\right) \left(1 + \frac{b_2}{u_2}\right)}{u_1 } \right], \\
	& \sigma^2_{\rm res} = \left(12 u_1^2 u_2^2\right)^{-1}\Big[ 3 \left(2+r b_1\right){}^2 b_2^2+(3 \left(2+r b_1\right){}^2+2 \left(r+2 r^3\right) b_1 u_1  +r^2 \left(1+2 r^2\right) u_1^2) u_2^2 \\& +2 b_2 \left(3 \left(2+r b_1\right){}^2 u_2+u_1 \left(12+r b_1 \left(6+u_2+2 r^2 u_2\right)\right)\right) \Big],\\
w=3:&  \\
	& \mu_{\rm res} = \frac{1}{2} \left[ r^2 +  \frac{\left(2 + r b_1 \right) \left(1+\frac{b_2 \left(b_3+u_3\right)}{u_2 u_3}\right)}{u_1} \right], \\
	& \sigma^2_{\rm res} = \left( 12 u_1^2 u_2^2 u_3^2 \right)^{-1} \Big[\left(3 \left(2+r b_1\right){}^2+2 \left(r+2 r^3\right) b_1 u_1+r^2 \left(1+2 r^2\right) u_1^2\right) u_2^2 u_3^2 
	+3 \left(2+r b_1\right){}^2 b_2^2 \left(b_3+u_3\right){}^2 \\
	&+2 b_2 \Big(6 \left(2+r b_1\right) b_3 u_1 \left(b_3+u_2\right) +b_3 \left(3 \left(2+r b_1\right){}^2 u_2+u_1 \left(24+r b_1 \left(12+u_2+2 r^2 u_2\right)\right)\right) u_3 \\ 
	&+\left(3 \left(2+r b_1\right){}^2 u_2+u_1 \left(12+r b_1 \left(6+u_2+2 r^2 u_2\right)\right)\right) u_3^2\Big) \Big],
\end{align*}
where $r$ is considered dimensionless, an integer corresponding to the number of lattice sites from the target site.

\section{Numerical inverse Laplace transform}
\label{sec:inverse-laplace}
Based on previous experience, we pick the Gaver-Stehfest method~\cite{gaver_observing_1966, stehfest_algorithm_1970} for the numerical inversion of Laplace transforms, which only requires the evaluation of the transformed function at real values of the Laplace variable and is suitable for bounded functions such as first-passage time distributions. We employ the algorithm described by Abate and Whitt~\cite{abate_unified_2006} to approximate the inverse Laplace transform of $\widetilde{f}(\epsilon)$ as
\begin{align*}
f(t) &= \frac{\ln 2}{t} \sum_{k=1}^{2M} w_k \widetilde{f}(k \frac{\ln 2}{t}), \\
w_k &= (-1)^{M+k} \sum_{j=\left\lfloor (k+1)/2 \right\rfloor }^{\min (k,M)} j^{M+1} \frac{2j!}{(j!)^2(M-j)!(2j-k)!(k-j)!}, \nonumber
\end{align*} 
where $M$ is a positive integer and $\left\lfloor (k+1)/2 \right\rfloor$ means the largest integer less than or equal to $(k+1)/2$. The number $M$ is chosen based on the available numerical precision, and according to the estimate provided in ref.~\cite{abate_unified_2006}, the result has around $2.2M$ digits of precision.

\section{Comparison with simulations}
\label{sec:simulation}
Assembly formation process is simulated according to the model description given in Section \ref{sec:model-description}, using the Gillespie algorithm~\cite{gillespie_exact_1977} that proceeds by determining the time of the next transition in the system, as well as which transition is going to take place. The system state is then updated, and the process is repeated until the random walker: 1) arrives at the bound state $w$, 2) decays (at rate $\gamma$ while at an unbound state), or 3) reaches one of the sites $-r$ or $r$, for the computation of $T_{\rm comp}$, $T_{\rm dis}$ and $T_{\rm res}$, respectively. $P_{\rm comp}$ is obtained by computing the ratio of the number of times the assembly forms to the total number of independent simulation runs. In each simulation run, the system starts at the same state where the random walker is occupying the first bound state with certainty, in accordance with the model description in Section \ref{sec:model-description}. 
\begin{table}[h]
\centering
\begin{tabular}{|c|c|c|c|c|c|}
\hline
$u_*$ & $w$ & $P_{\rm comp}^{\rm seq}$ [Eq. \eqref{eqn:Pcomp}] & $P_{\rm comp}^{\rm seq}$ (sim) & $P_{\rm comp}^{\rm ran}$ [Eq. \eqref{eqn:Pcomp}] & $P_{\rm comp}^{\rm ran}$ (sim)\\ \hline
0.10& 2& 0.5076& 0.5078 [0.5071, 0.5086]& 0.5076& 0.5084 [0.5076, 0.5092]\\ \hline
0.40& 3& 0.2839& 0.2840 [0.2832, 0.2847]& 0.4423& 0.4420 [0.4413, 0.4429]\\ \hline
0.08& 4& 0.4202& 0.4206 [0.4199, 0.4213]& 0.7101& 0.7105 [0.7099, 0.7111]\\ \hline
0.19& 5& 0.2708& 0.2699 [0.2694, 0.2705]& 0.6863& 0.6859 [0.6852, 0.6866]\\ \hline
0.13& 6& 0.3397& 0.3398 [0.3394, 0.3402]& 0.8007& 0.8009 [0.8005, 0.8014]\\ \hline
0.31& 7& 0.0858& 0.0858 [0.0855, 0.0861]& 0.5531& 0.5527 [0.5520, 0.5534]\\ \hline
0.15& 8& 0.2979& 0.2983 [0.2976, 0.2990]& 0.8571& 0.8570 [0.8566, 0.8575]\\ \hline
\end{tabular}
\caption{Comparison of numerically exact and simulated values of $P_{\rm comp}$. Values of $u_*$ and $w$ are indicated in the first two columns, and  all other parameter values are fixed at: $f=1$, $b_1=2$, $u_1=1$, $b_*=0.25$, $\gamma=0.1$. In computing simulation results, an ensemble of $N=10^5$ independent simulation runs were obtained, and this process was repeated 20 times to compute error due to finite sample size. $P_{\rm comp}^{\rm seq}$ (sim) and $P_{\rm comp}^{\rm ran}$ (sim) correspond to simulation results, given in the format x [y, z], where $x$ is the average, and [y, z] is the 95\% confidence interval computed via bootstrapping.}
\label{table:Pcomp}
\end{table}

In Table \ref{table:Pcomp} and Fig. \ref{fig:comparision_with_MC}, we display the comparison between analytical and simulation results for $P_{\rm comp}$ and the cumulative distributions of $T_{\rm comp}$, $T_{\rm dis}$ and $T_{\rm res}$, for several randomly selected parameter values. The cumulative probability distribution of $T_x$, which is straightforward to calculate from simulation data without any binning, is defined as
\begin{gather}
F_{x}^{y}(t) = \int_0^t ds\, f_x^y(s) = \mathcal{L}^{-1}\left\{\frac{\widetilde{f}_x^y(\epsilon)}{\epsilon}\right\},
\label{eqn:CDF_formal}
\end{gather}
where y is either ``seq" or ``ran", for sequential and random binding models, respectively, and the inverse Laplace transform, denoted by $\mathcal{L}^{-1}$, is performed as in Appendix \ref{sec:inverse-laplace}. Parameter values are given in captions. As both sets of comparisons show, numerically exact and simulation results that are obtained independently are in excellent agreement.

\begin{figure}[h]
\centering
\includegraphics[width=0.95\columnwidth]{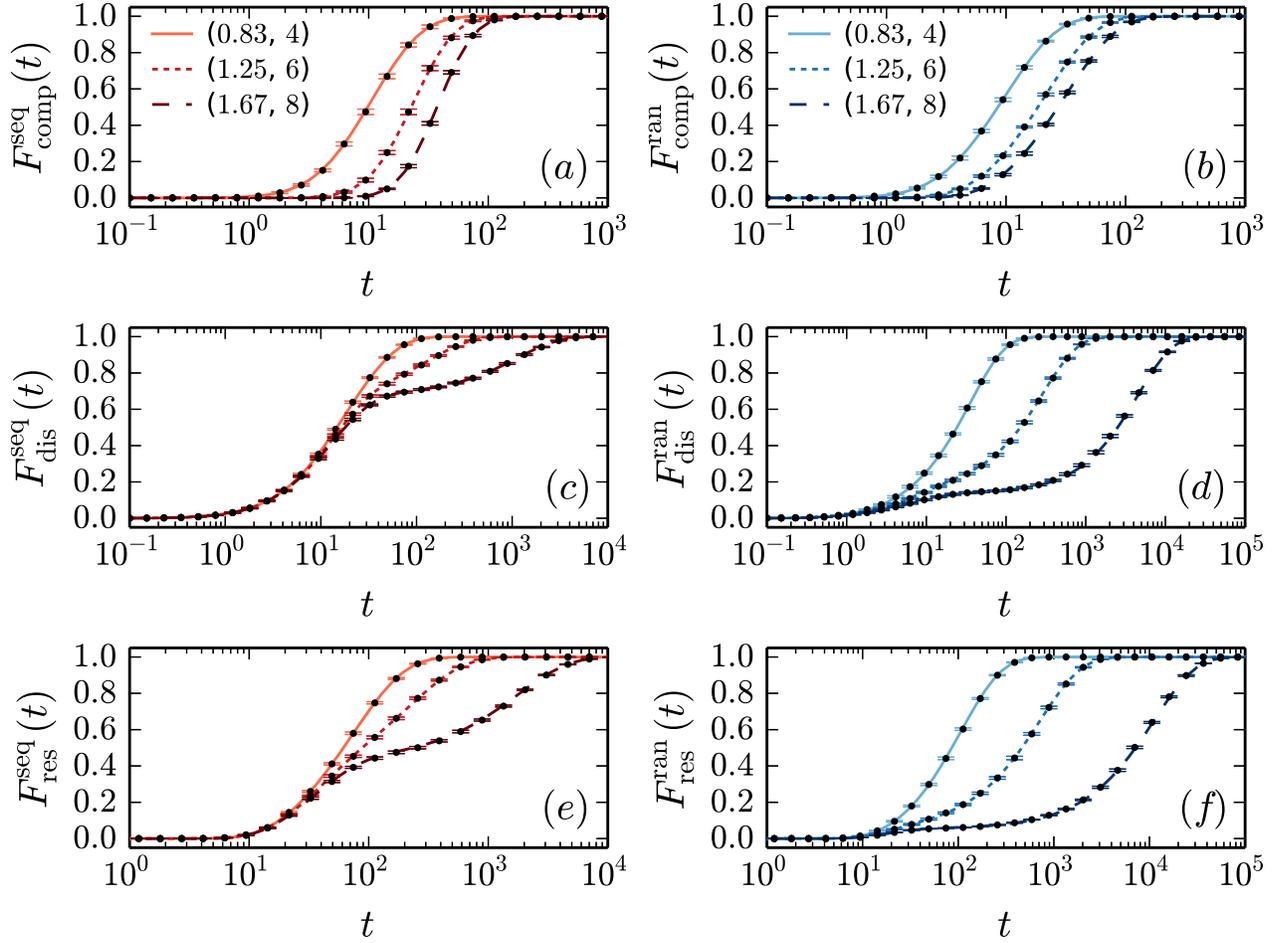}
\caption{(Color online) Comparison of numerically exact and simulated cumulative probability distributions of $T_{\rm comp}$, $T_{\rm dis}$ and $T_{\rm res}$, defined in \eqref{eqn:CDF_formal}. Curves correspond to the numerical inverse Laplace transform of \eqref{eqn:CDF_formal} (with the use of Eqs. \eqref{eqn:f-comp-laplace-appendix}, \eqref{eqn:f-dis-laplace-explicit} and \eqref{eqn:f-res-laplace}), and black filled circles along with error bars represent simulation results. Legends of the topmost graphs apply to all graphs in the same column, and values in parenthesis correspond to $(b_*/u_*,\,w)$. For the calculation of $F_{\rm res}$, we set $r=10$. Rest of the parameter values are the same as those in the caption of Table. \ref{table:Pcomp}. In computing simulation results, an ensemble of $N=10^3$ independent simulation runs were obtained, and this process was repeated 20 times to compute error bars. Error bars correspond to 95\% confidence intervals computed via bootstrapping.}
\label{fig:comparision_with_MC}
\end{figure}

\end{widetext}

\newpage

\bibliography{./transcription_complex}

\end{document}